\pdfoutput=1
\documentclass[aps,
pra,
twocolumn,
superscriptaddress,
longbibliography,
footinbib,
]{revtex4-1} 

\usepackage{amsmath,amssymb,amstext}
\usepackage{mathtools}
\usepackage{braket}
\usepackage{xcolor}
\usepackage[utf8]{inputenc} 
\usepackage{graphics}
\usepackage{bm}
\usepackage[colorlinks=true,citecolor=blue,linkcolor=blue,urlcolor=blue]{hyperref}
\usepackage[normalem]{ulem}
\usepackage{mathrsfs}

\usepackage[mathlines]{lineno}

\newcommand{\diff}[1]{\mathrm{d}#1\,} 
\newcommand{\average}[1]{\overline{#1}}
\DeclarePairedDelimiter{\abs}{\lvert}{\rvert}
\DeclarePairedDelimiter{\expval}{\langle}{\rangle}

\let\Im\relax
\DeclareMathOperator{\Im}{Im}
\let\Re\relax
\DeclareMathOperator{\Re}{Re}
\def\be{\begin{equation}}
\def\ee{\end{equation}}

\begin{document}

\title{Quantum boomerang effect for interacting particles}

\author{Jakub Janarek}
    \email{jakub.janarek@lkb.upmc.fr}
    \affiliation{Institute of Theoretical Physics, Jagiellonian University in Krakow, \L{}ojasiewicza 11, 30-348, Krak\'ow, Poland}
   \affiliation{Laboratoire Kastler Brossel, Sorbonne Universit\'e, CNRS, ENS-Universit\'e PSL, Coll\`ege de France; 4 Place Jussieu, 75004 Paris, France }
     \author{Dominique Delande}%
     \email{dominique.delande@lkb.upmc.fr}
\author{Nicolas Cherroret}
     \email{nicolas.cherroret@lkb.upmc.fr}
    \affiliation{Laboratoire Kastler Brossel, Sorbonne Universit\'e, CNRS, ENS-Universit\'e PSL, Coll\`ege de France; 4 Place Jussieu, 75004 Paris, France }
\author{Jakub Zakrzewski}
    \email{jakub.zakrzewski@uj.edu.pl}
    \affiliation{Institute of Theoretical Physics, Jagiellonian University in Krakow, \L{}ojasiewicza 11, 30-348, Krak\'ow, Poland}
    \affiliation{Mark Kac Complex Systems Centre, Jagiellonian University in Krakow, \L{}ojasiewicza 11, 30-348, Krak\'ow, Poland}%

\date{\today}

\begin{abstract}
When a quantum particle is launched with a finite velocity in a disordered potential, it may surprisingly come back to its initial position at long times and remain there forever. This phenomenon, dubbed  ``quantum boomerang effect'', was introduced in [Phys. Rev. A {\bf 99}, 023629 (2019)]. Interactions
between particles, treated within the mean-field approximation, are shown to partially destroy the boomerang effect: the center
of mass of the wave packet makes a U-turn, but does not completely come back to its initial position. We show 
that this phenomenon can be quantitatively interpreted using a single parameter, the average interaction energy.  
\end{abstract}

\keywords{Disordered systems, Anderson localization, interacting particles, Bose-Einstein condensation}
\maketitle


\section{Introduction}
Anderson Localization (AL) \cite{anderson1958absence}, i.e. inhibition of transport in disordered media, has been the source of various, often counter-intuitive phenomena discovered over the last 60 years \cite{Evers08}.  Already in one dimension (1D) the fact that even the tiniest random disorder generically leads to a full localization of eigenstates is totally against a classical way of thinking. This effect is a clear manifestation of the inherently interferometric nature of AL, typically explained as the effect of quantum wave interference. The phenomenon was observed in many experimental setups including light \cite{chabanov2000statistical, schwartz2007transport}, sound waves \cite{hu2008localization} as well as matter waves \cite{billy2008direct, jendrzejewski2012three, manai2015experimental, semeghini2015measurement, white2019observation}. 
The closely related phenomenon of Aubry-Andr\'e localization in quasi-periodic potentials \cite{Aubry80} was also observed in a cold atomic setting \cite{Roati08}.  
Last years have also led to a number of studies of the many-body counterpart of AL \cite{Gornyi05, basko2006metal}, the so called many-body localization (MBL)  (for recent reviews see \cite{Huse14,Alet18}). It has already been  observed in cold atomic experiments in quasi-periodic potentials \cite{schreiber2015observation, choi2016exploring}. While studies of MBL have been very extensive, even its very existence has been questioned recently \cite{Suntajs19}, provoking a vivid debate \cite{Abanin19z,Sierant19z,Panda19}. 

The physics of a single particle in a random potential, particularly in 1D, has much stronger foundations, while still bringing novel features such as studies of random (or quasi-random) potentials revealing the presence of mobility edge \cite{Luschen18,Major18}, i.e. situations where localized and extended states appear at different energies.   Even for a pure random standard case, one may find new counter-intuitive  phenomena as exemplified by the \emph{quantum boomerang effect}~\cite{prat2019quantumboomerang}. As a classical boomerang returns to the initial position from which it was launched, the center of mass of a wave packet with a nonzero initial velocity returns to its origin due to AL. The effect is quite general and occurs
in Anderson localized multi-dimensional systems~\cite{prat2019quantumboomerang} whose Hamiltonians preserve time-reversal invariance.

The aim of this work is to investigate how interactions between particles affect the boomerang effect.
Our study 
focuses on the limit of weak interactions, a regime where it was previously shown that AL of wave packets is replaced by
 a subdiffusive evolution at very long time \cite{pikovsky2008destruction, skokos2009delocalization, flach2009universal, Cherroret14, vakulchyk2018universal} - see however \cite{Sacha09,delande2013manybody}. 
Computing the temporal evolution of a many-body wave packet in a disordered potential is in general a formidable task, even in 1D.
When interactions are sufficiently weak however, one may use a mean-field approximation and describe the dynamics with a one-dimensional Gross-Pitaevskii equation (GPE). Within this formalism, we provide an in depth numerical analysis of the quantum boomerang effect. For simplicity,
we study the weak-disorder case, for which an analytic description of the boomerang effect in the absence of interaction is available. 
At short times, of the order of a few disorder scattering times, we observe that the dynamics of the wave packet is essentially not affected by 
interactions, with an initial ballistic flight followed by a U-turn of the center of mass, slowly returning towards its initial position. 
The main effect in this regime is a small modification of the scattering time and scattering mean free path due to interactions. At longer time, we observe that the center of mass, instead of returning to its initial position, stops at a finite distance from it,
which increases with the interaction strength.
We show that this phenomenon can be interpreted in terms of a break time, the time scale beyond which interference effects are typically destroyed by interactions. We finally show that in the presence of interactions, the boomerang effect can be quantitatively described in terms of a single scaling parameter, a variant of the interaction energy, computed at the break time.

The paper is organized as follows. After formulating the problem and introducing the main parameters in Sec. \ref{sec:model}, we numerically analyze the influence of interactions  on the boomerang effect in Sec. \ref{sec:lessening}. In Sec. \ref{sec:universal scaling}, we then perform several numerical studies which establish that the dynamics can be described by a single parameter. The appendix shows that the short-time dynamics can also be understood using the same parameter.   
We finally conclude and briefly discuss open questions.

\section{The model}
\label{sec:model}

To study the boomerang effect in the presence of interactions, we use the one-dimensional, time dependent Gross-Pitaevskii equation (GPE) \cite{pitaevskii2016bose}:
\begin{equation}\label{eq:gpe}
    i\hbar\frac{\partial\psi(x,t)}{\partial t} = \left[\frac{p^2}{2m} + V(x)+ g\abs{\psi(x,t)}^2\right]\psi(x,t),
\end{equation}
where $V(x)$ is a disordered potential, while $g$ represents the interaction strength. Wave functions are normalized to unity, $\int\abs{\psi(x,t)}^2\diff{x}=1$. Throughout this work we assume that the disordered potential is a Gaussian uncorrelated random variable, i.e. 
\begin{equation}
\average{V(x)} = 0, \ \ \ \ \ \average{V(x)V(x')} = \gamma\delta(x-x'),
\end{equation}
where the overbar denotes the average over disorder realizations. 
The parameter $\gamma$ measures
the disorder strength and determines the characteristic time and length scales 
for scattering: the mean scattering time $\tau_0$ (the typical time for a
particle to be scattered by the disorder) and the corresponding mean scattering length $\ell_0$, usually called mean free path. 
The uncorrelated disorder model is sufficient to capture the main features of the boomerang effect, which depends only on a small set of well defined parameters, $\tau_0$, $\ell_0$
and time $t$ itself. In the Born approximation, $\tau_0$ and $\ell_0$ are given by~\cite{akkermans2007mesoscopic}:
\begin{equation} 
    \tau_0 = \frac{\hbar^3k_0}{2m
    \gamma},\quad \ell_0 = \frac{\hbar k_0}{m}\tau_0 = \frac{\hbar^4 k_0^2}{2m^2\gamma}.
 \label{eq:born}
\end{equation}
Non-interacting 1D systems remain always strongly localized, i.e. their eigenstates decay exponentially over the localization length $\xi_\text{loc} = 2\ell_0$ \cite{lifshits1988introduction}. 

Following \cite{prat2019quantumboomerang}, as the initial state for time evolution we  take a Gaussian wave packet with mean velocity $\hbar k_0/m$:
\begin{equation}
\psi(x,t\!=\!0) = \left(\frac{1}{\pi\sigma^2}\right)^{1/4} \exp(-x^2/2\sigma^2 + i k_0 x),
\end{equation} 
where $\sigma$ and $k_0$ are chosen such that the initial wave function is sharply peaked around $\hbar k_0$ in momentum space, i.e. $k_0\sigma \gg 1$. 
Moreover, we focus on the weak-disorder regime where the mean free path is much longer than the de Broglie wavelength, that is $k_0\ell_0 \gg 1$ \cite{akkermans2007mesoscopic}. In this regime, a full analytic theory, based on the Berezinskii diagrammatic technique~\cite{berezinskii1974kinetics}
exists in the non-interacting limit. It makes it possible to express the average center-of-mass position 
(CMP):
 \begin{equation}\label{eq:com_wave_packet}
 \expval{x(t)} = \int x\average{\abs{\psi(x,t)}^2} \diff{x},
 \end{equation}
as:
\begin{equation}
\frac{\expval{x(t)}}{\ell_0} = f\left(\frac{t}{\tau_0}\right),
\label{eq:x_vs_t}
\end{equation}
where $f$ is a universal function whose Taylor expansion at short time and asymptotic behavior at long time are exactly known~\cite{prat2019quantumboomerang}. In particular $\expval{x(t)}\approx t\ell_0/\tau_0$ at short time $t\ll \tau_0$, meaning that the initial motion is ballistic, and
\begin{equation}
\expval{x(t)} \approx 64\ell_0 \frac{\ln(t/4\tau_0)\tau_0^2}{t^2}
\label{eq:analytic_x}
\end{equation}
at long time $t\gg \tau_0.$
The two assumptions of weak disorder, $k_0\ell_0\gg 1,$ and narrow wave packet in momentum space, $k_0\sigma\gg 1,$ make it possible to have a well controlled non-interacting limit~\footnote{Note that there is no constraint on the size of the wave packet $\sigma$ compared to the mean free path $\ell$. The expression
(\ref{eq:x_vs_t}) does not depend on the ratio of the these two length scales.}. If they are not valid, the wave packet will contain
a broad energy distribution, and consequently a distribution of scattering time
(which depends on energy), making the analysis more tedious. However, the phenomena described below are expected to be very similar.

Another important ingredient of the model are interactions. Our interest is only in small values of $g$, which corresponds to a weak interaction regime \cite{cherroret2011fokker, geiger2013microscopic, wellens2019private}.

In the following, we study numerically the propagation of the initial wave function $\psi(x,t=0)$ for different disorder realizations. It yields the averaged density profile $\average{\abs{\psi(x,t)}^2}$, from which we compute the CMP,
Eq.~(\ref{eq:com_wave_packet}).
The numerical technique is as follows. The one-dimensional configuration space is discretized on a regular grid, over which the wave function is computed. The temporal propagation is performed using a split-step algorithm of step $\Delta t$, alternating
propagations of the linear part of the GPE, $\exp\left[-i(p^2/2m+V)\Delta t/\hbar\right]$, and of the nonlinear part, which is simply
a phase factor $\exp(-i g \abs{\psi(x,t)}^2\Delta t/\hbar).$ The linear part of the evolution operator is expanded in a series of Chebyshev polynomials, as described in  \cite{Tal-Ezer84, Cheby91, roche1997conductivity,fehske2009numerical}.  

Throughout our work, we express lengths in units of $1/k_0$, times in $m/\hbar k_0^2$, and energies in $\hbar^2k_0^2/m$. The interaction strength is expressed in units of $\hbar^2k_0/m$, and the disorder strength $\gamma$ in units of  $\hbar^4k_0^3/m^2.$ 

Numerical results have been obtained on a discretized grid of size $4000/k_0$ (sufficiently large for the wave function to be vanishingly small at the edges; open boundary conditions have been used) divided into 20000 points, so that discretization effects are negligible.
	 We have used a disorder strength $\gamma = 0.1 \hbar^4k_0^3/m^2,$ so that, in the Born approximation (\ref{eq:born}), one has 
	 $k_0\ell_0=5,$ i.e. the disorder is weak. 
	 Note that, while the true mean free path and scattering time may slightly differ 
	 from their expressions~(\ref{eq:born}) at the lowest order Born approximation, the dynamics of the non-interacting quantum boomerang effect is still given by Eq.~(\ref{eq:x_vs_t}), provided corrected values of $\tau_0$ and $\ell_0$ are used. We have performed calculations for various widths $\sigma$ of the initial wave packet $\psi(x,t=0),$ as well as various values of the interaction strength $g$.

\section{The boomerang effect with interactions}\label{sec:lessening}

\subsection{Role of the interaction strength}

\begin{figure}
    \centering
    \includegraphics[width=\columnwidth]{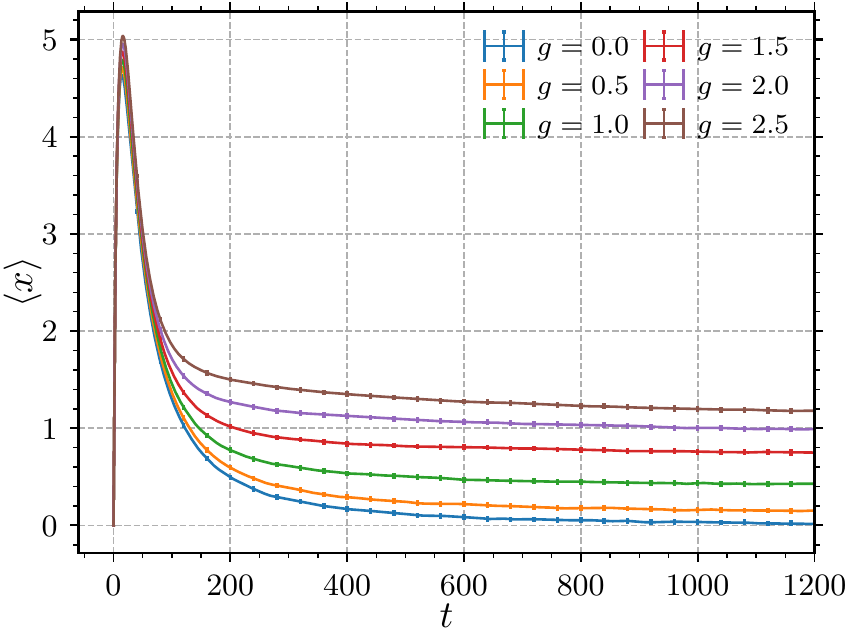}
    \caption{
    CMP time evolution for an initial wave packet with $\sigma=10/k_0$, for different values of the interaction strength $g$. Here, $\expval{x}$ is in units of $1/k_0$ and time $t$ in units of $m/\hbar k_0^2$. The legend indicates curves from bottom to top.  All curves have been averaged over more than $5\times10^5$ disorder realizations.  The short-time behavior  remains almost unchanged, whereas the long-time evolution clearly depends on the interaction strength. The error bars represent statistical average errors. Center-of-mass trajectories among different disorder realizations are normally distributed, such that we use the standard error of the mean as estimator of the errors.
    }
    \label{fig:com_sigma_10_tmax1200}
\end{figure}

\begin{figure}
 \centering
 \includegraphics[width=\columnwidth]{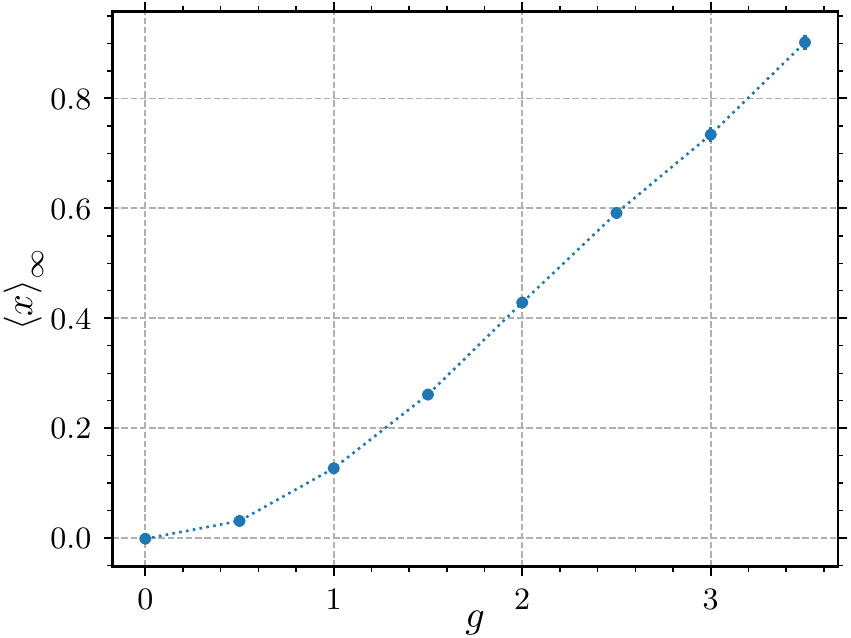}
 \caption{Long-time average $\expval{x}_\infty$ vs. interaction strength $g$ for a wave packet with initial width $\sigma=40/k_0$. $\expval{x}_\infty$ is in units of $1/k_0$, and $g$ is in units of $\hbar^2 k_0/m$. Data have been averaged over more than $5\times10^5$ disorder realizations. }
 \label{fig:longtime_x_vs_g_sigma40}
\end{figure}

In Fig. \ref{fig:com_sigma_10_tmax1200} we present results obtained for a wave packet with initial width $\sigma=10/k_0$ for various interaction strengths $g$. Similarly to the non-interacting case, after the initial ballistic motion of the center of mass, we observe a subsequent reflection towards the origin, that is a boomerang effect. However, the long-time behavior is affected by interactions.  For non-zero values of $g,$ the center of mass does not return to the origin but saturates at some finite value. In other words, the boomerang effect is only partial~\footnote{A preliminary study of this partial destruction of the boomerang effect can be found on the arXiv preprint~\cite{prat2017anderson}. However, these preliminary results are not part of the published paper~\cite{prat2019quantumboomerang}}. This can be understood as follows. For $g=0,$ the disorder is static and full Anderson localization sets in; the complicated interference between multiply scattered paths leads to full localization at infinite time, and also to full return of the center of mass to its initial position. For non-zero but small $g$, the nonlinear term in Eq.~(\ref{eq:gpe}), $g|\psi(r,t)|^2$, plays the role of a small additional effective potential which is \emph{time-dependent}, thereby adding a fluctuating phase along each scattering path. This breaks the interference between multiple scattering paths and thus destroys both Anderson localization~\cite{pikovsky2008destruction, skokos2009delocalization, flach2009universal, Cherroret14,vakulchyk2018universal} and the full boomerang effect at long time.  
For all studied widths of the wave packet, namely $k_0\sigma=5,\, 10,\, 20$ and $40$ we observe a similar saturation effect.
Of course, the phase scrambling progressively develops over time. The characteristic break time over which it kills coherent transport and boomerang effect is discussed in detail in the sequel of this paper. Note that the interpretation of the effect of interactions in terms of a decoherence mechanism may be questioned at very long time, where thermalization comes into play and affects the momentum distribution as well as the dynamics of the system \cite{Tavora2014, Miniatura15}. In our case, both disorder and interactions are small, so that thermalization takes place at times significantly longer than the ones considered here.
Neglecting thermalization nevertheless restricts our analysis to the regime where the long-time CMP $\expval{x}_\infty\ll \ell_0$ (see Eq. (\ref{eq:longtime_average}) below), which constrains the maximum value of $g$. 

To study in detail the long-time evolution, we run numerical simulations up to time $t_\text{max}\approx2500\tau_0$. From these simulations we calculate the long-time average of the
CMP,
$\expval{x}_\infty$, defined in the following way:
\begin{equation}\label{eq:longtime_average}
 \expval{x}_\infty = \frac{1}{t_2-t_1} \int_{t_1}^{t_2}\expval{x(t)}\diff{t},
\end{equation}
where we choose $t_1\approx1200\tau_0$, $t_2\approx2500\tau_0$. The results are essentially independent 
of these bounds, provided they are much longer than $\tau_0$. This definition provides us with a very good estimate of the infinite-time 
CMP. Figure \ref{fig:longtime_x_vs_g_sigma40} shows the dependence of  $\expval{x}_\infty$ on the interaction strength~$g$. 
For small values of  $g,$ the CMP dependence is quadratic in $g$, $\expval{x}_\infty \propto g^2$, and becomes approximately linear for larger $g$. In the following, we will mostly concentrate on the quadratic regime of small interactions, leaving the more difficult case of larger $g$ for future studies.

	\begin{figure}
		\centering
		\includegraphics[width=\columnwidth]{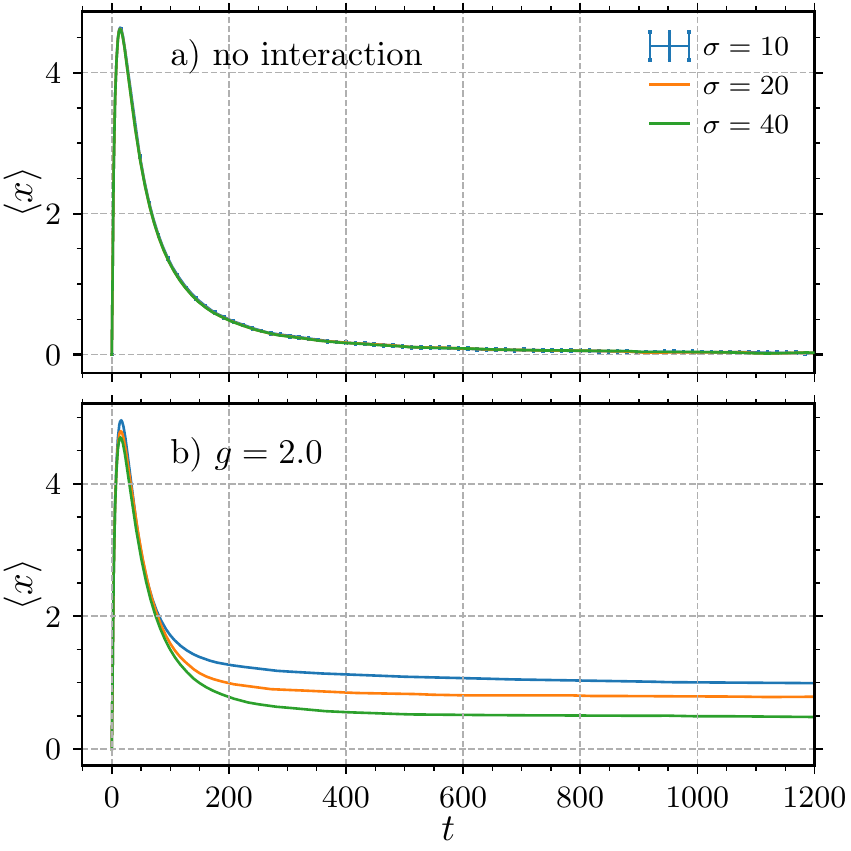}
		\caption{
		a) Comparison of non-interacting CMP $\expval{x(t)}$ for wave packets of widths $\sigma=10/k_0,\, 20/k_0,\,40/k_0$. All three curves 
		overlap, indicating that $\expval{x(t)}$ is $\sigma$-independent in the non-interacting limit.
		b) 
		Same as a) but 
		for non-zero interaction strength $g=2.0 \hbar^2 k_0/m$ (the legend indicates curves from top to bottom). Here, the saturation point of $\expval{x(t)}$ is higher for initially narrower wave packets. The center of mass $\expval{x(t)}$ is in units of $1/k_0$ and time $t$ in units of $m/\hbar k_0^2$.  Error bars are only shown in panel a) for
		$\sigma=10/k_0$ to indicate their order of magnitude. }
		\label{fig:com_vs_time_free}
	\end{figure}

\subsection{Role of the wave-packet width}

Another important parameter is the width $\sigma$ of the initial wave packet. In Fig. \ref{fig:com_vs_time_free}, we show $\expval{x(t)}$ for different $\sigma$ values.
While, for $g=0,$ it follows the analytic prediction (\ref{eq:analytic_x}) \emph{independently} of $\sigma,$ for interacting particles the long-time behavior strongly depends on $\sigma.$ A simple qualitative explanation is that the destruction of the boomerang effect is controlled by the 
nonlinear term $g\abs{\psi(x)}^2$ in Eq. (\ref{eq:gpe}). This term is larger for a spatially narrow wave packet,
so that interference between scattered waves are suppressed at shorter time, giving a higher $\expval{x}_\infty$ value.

\begin{figure}
		\centering
		\includegraphics[width=\columnwidth]{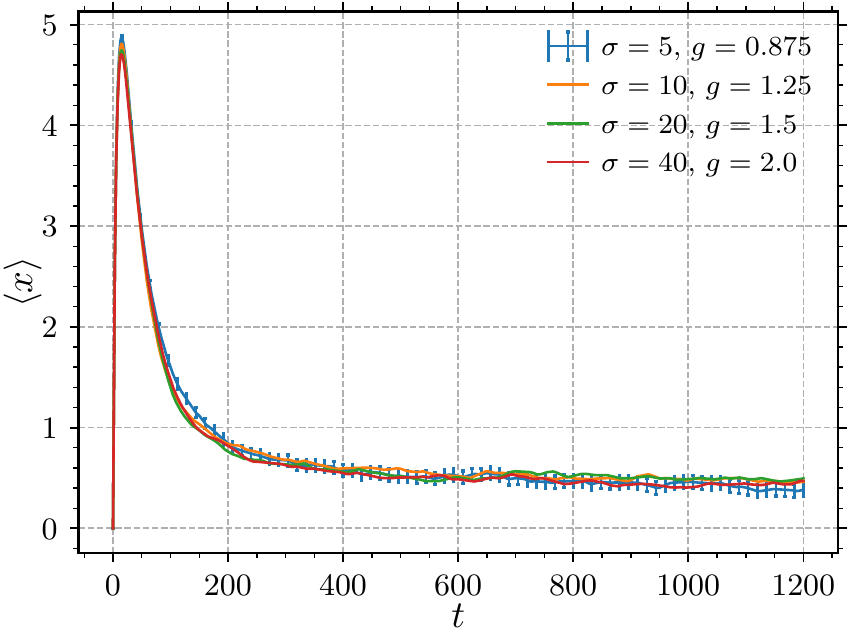}
		\caption{
		CMP $\expval{x(t)}$ vs. time for different initial states chosen such that all of them saturate around the value $\overline{\expval{x}}=0.4/k_0$. All the curves overlap. Results have been averaged over 16000 disorder realizations. {
		The CMP is shown in units of $1/k_0$ and time is in units of $m/\hbar k_0^2$.} Error bars are shown only for $\sigma=5/k_0$.
		}\label{fig:com_sigma_final_x}
	\end{figure}

Although the boomerang effect is affected by a change of either $g$ or $\sigma$ in the interacting case, one may guess that the CMP is not a function of these two independent parameters. Indeed, closer investigation reveals  that similar ``trajectories'' of $\expval{x(t)}$ can be achieved for different combinations of $g$ and  $\sigma$.
In particular, the same values of $\expval{x}_\infty$ can be obtained  from different initial states, provided $g$ is properly adjusted. 
This property is illustrated in Fig.~\ref{fig:com_sigma_final_x}, where we have computed $\expval{x(t)}$ for various values of $\sigma$ and have adjusted $g$ so to that the curves fall on top of each other.

\section{Universal scaling of the interacting boomerang effect}
\label{sec:universal scaling}

From the results of the previous section, it is natural to ask whether or not the interacting boomerang effect, and more specifically its long-time average $\expval{x}_\infty$ -- Eq. (\ref{eq:longtime_average}) --  can be described in terms of a \textit{single} parameter, in the spirit of scaling approaches well-known in the context of Anderson localization of non-interacting particles \cite{Abrahams79}.

\subsection{Break time}

Before attempting to rescale the CMP, let us introduce an important parameter that will turn useful in the following. We recall that in the non-interacting limit, the CMP at long time is given by Eq.~(\ref{eq:analytic_x}). If we neglect the logarithmic part, $\expval{x(t)}$ decays as $t^{-2}$.  It suggests that one may identify  a characteristic time scale connected with  weak interactions  which is inversely proportional to $g$. In our analysis we call this time scale the \emph{break time} $t_b$ and define it by the relation:
\begin{equation}\label{eq:breaktime}
 \expval{x(t_b)}_{g=0} = \expval{x}_\infty(g),
\end{equation}
where for the left-hand-side we use the analytical prediction of the non-interacting theory, Eq. (\ref{eq:analytic_x}). A given time scale in quantum mechanics corresponds to some energy scale, here the \emph{break energy}:
\begin{equation}\label{eq:breakenergy}
E_b = \frac{2\pi\hbar}{t_b}, 
\end{equation} 
which will turn out to be a key parameter in our rescaling of the CMP.

\subsection{Rescaling of the CMP: first attempt}

\begin{figure}
\centering
 \includegraphics[width=\columnwidth]{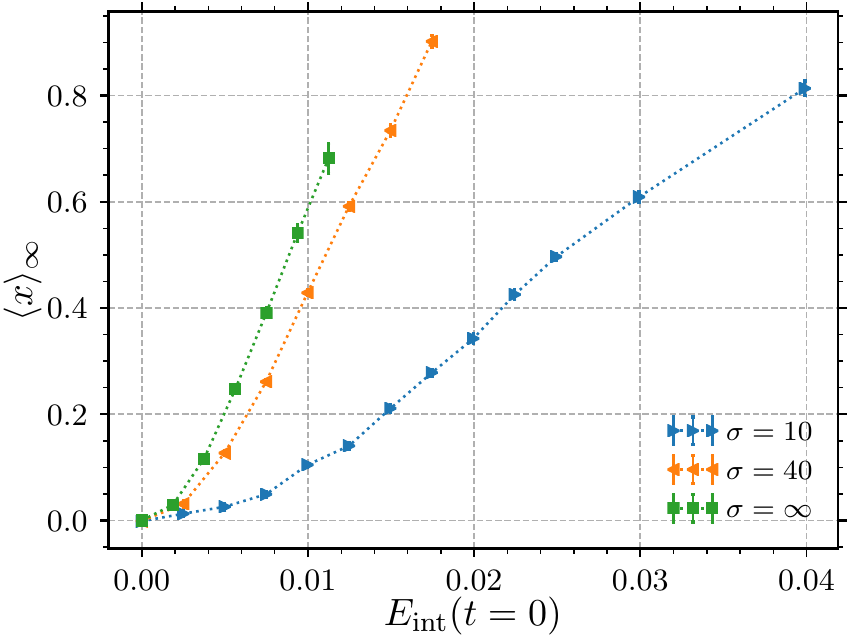}
 \caption{ Dependence of the long-time average $\expval{x}_\infty$, Eq. (\ref{eq:longtime_average}), on the initial interaction energy. For wave packets $E_\text{int}(t~=~0) = g/(2\sqrt{2\pi}\sigma)$, while $E_\text{int}(t=0) = g\rho_0/2$ in the plane-wave limit $\sigma=\infty$. The interaction energy is expressed in units of $\hbar^2 k_0^2/m$, and the CMP in units of $1/k_0$. Error bars represent standard deviation of the averaged points.}\label{fig:longtime_x_vs_g_normalized}
\end{figure}
\begin{figure}
 \centering
 \includegraphics[width=\columnwidth]{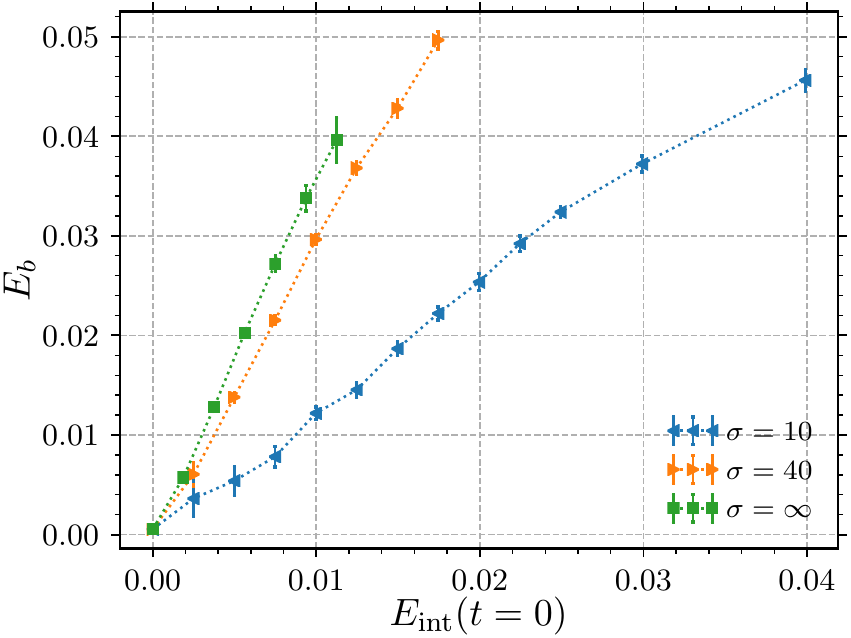}
 \caption{Break energy $E_b$ vs. initial interaction energy $E_\text{int}(t=0)$. The break energy is defined by $E_b = 2\pi\hbar/t_b$, where the break time $t_b,$ calculated according to Eq. (\ref{eq:breaktime}), is the characteristic time beyond which the boomerang effect disappears. Energies are in units of $\hbar^2k_0^2/m$. Error bars on break energy points represent error on break time calculation from long time average of CMP. 
 }\label{fig:breakenergy_vs_initial_int_energy}
\end{figure}

A closer inspection of Fig. \ref{fig:com_sigma_final_x} reveals that, when trying to superimpose the CMP curves, broader wave packets require higher interaction strengths.
A first natural candidate to  characterize the CMP is thus the interaction energy,
\begin{equation}\label{eq:interaction_energy}
    E_\text{int}(t) = \frac{g}{2}\int\average{\abs{\psi(x,t)}^4} \diff{x}.
\end{equation}
We recall that the total energy, conserved by the GPE,  is the sum of the non-interacting part and the interaction energy: $E_\text{tot}=\expval{p^2}/2m + \expval{V} + E_\text{int}$.
A related important quantity is the interaction energy at initial time,  $E_\text{int}(t=0)=g/(2\sqrt{2\pi}\sigma)$. 
Notice that while the CMP curves in Fig. \ref{fig:com_sigma_final_x} are obtained from quite different values of $g$ and $\sigma$, they are all associated with comparable interaction energies at time $t=0$.
This is a very clear hint that $E_\text{int}(t=0)$ is a crucial parameter for describing the impact of interactions on the boomerang effect.
We can thus try to rescale the results by plotting them vs. the initial interaction energy. This is done for $\expval{x}_\infty$ in Fig. \ref{fig:longtime_x_vs_g_normalized}, and for the break energy -- see Eq. (\ref{eq:breakenergy}) -- in Fig. \ref{fig:breakenergy_vs_initial_int_energy}. 
In these plots we also show, for comparison, the formal limit $\sigma\to\infty$ of infinitely large wave packets,  where $\psi_0(x) \to \sqrt{\rho_0}\exp(ik_0x)$ reduces to a plane wave with $E_\text{int}(t=0)=g\rho_0/2$. In this limit, the sole parameter $g\rho_0$ controls the boomerang effect. Note that despite the flatness of the density profile in the plane-wave limit, in practice it is still possible to study the boomerang effect by measuring an effective CMP defined as: 
\begin{equation}\label{eq:com_plane_wave}
 \expval{x(t)}_{\sigma=\infty} \equiv \frac{1}{m}\int_0^t\expval{p(t')}\diff{t'}.
\end{equation}
We have verified numerically, in particular, that in the non-interacting case the definition (\ref{eq:com_plane_wave}) agrees with  results for wave packets of finite width, thus with the theoretical prediction (\ref{eq:analytic_x}).

 The curves in Figs. \ref{fig:longtime_x_vs_g_normalized} and \ref{fig:breakenergy_vs_initial_int_energy}, which correspond to different values of $\sigma$, are qualitatively similar, despite the fact that the $g$ values are widely different. This suggests that the interaction energy is indeed an important parameter. Moreover, note that the break energy $E_b$ is comparable (within a factor 4) to the initial interaction energy. In particular, because $E_\text{int}(t=0)$ is proportional to the interaction strength $g$ for all initial states, at small values of $E_\text{int}(t=0)$ we see the expected linear dependence of $E_b$ with $E_\text{int}(t=0)$. 
Nevertheless, it is clear that plots based on $E_\text{int}(t=0)$ do no fall on the same universal curve: $\expval{x}_\infty$ and $E_b$ deviate from each other as $\sigma$ is changed, approaching the upper limit curve $\sigma=\infty$  as $\sigma$ increases (this limit is represented by green square symbols in Figs. \ref{fig:longtime_x_vs_g_normalized} and \ref{fig:breakenergy_vs_initial_int_energy}).

\subsection{Nonlinear energy at break time}
\label{subsec:nonlinear_energy}

\begin{figure}
	\centering
	\includegraphics[width=\columnwidth]{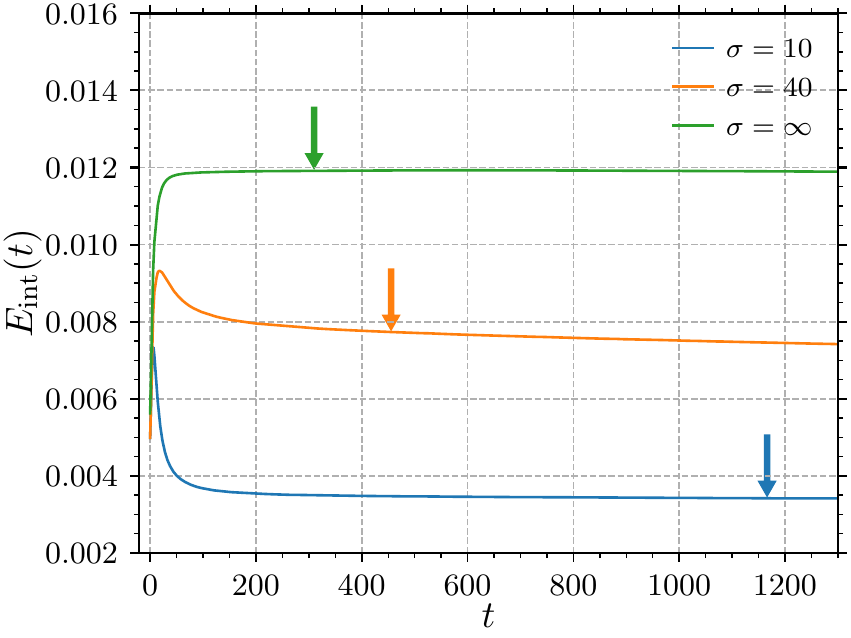}
	\caption{Interaction energy vs. time for different initial states and interaction strengths. The plot shows data for wave packets with $[\sigma=10/k_0, \,g=0.25\hbar^2k_0/m]$, $[\sigma=40/k_0, \,g=1.0\hbar^2k_0/m]$, and $[\sigma=\infty,\,g=45.0\hbar^2 k_0/m, \, \rho_0 = 0.00025k_0]$. $E_\text{int}$ is in units of $\hbar^2k_0^2/m$ and time $t$ in units of $m/\hbar k_0^2$. The legend indicates curves from bottom to top.
Initial states are chosen to clearly emphasize the randomization of the wave function amplitude at long time. It is, however, generically present for any interaction strength. 
In the plane-wave limit, randomization doubles $E_\text{int}$ over a short time scale comparable to the scattering time $\tau_0$, which then remains stationary
at long time. 
For finite $\sigma$, randomization is is visible at short time, but followed by a slow decay at long time, because of wave-packet spreading. From $t=t_b$ onwards, however, the decay is very slow (the location of $t_b$ is indicated by arrows).
}
	\label{fig:speckelization}
\end{figure}

The reason why $E_\text{int}(t=0)$ does not allow for a universal rescaling of the CMP stems from the fact that the interaction energy $E_\text{int}(t)$ varies significantly from $t=0$ onwards. This evolution is shown in Fig. \ref{fig:speckelization} for two finite values of $\sigma$ and for the plane-wave limit $\sigma=\infty$.

The figure reveals that the time evolution of the interaction energy generally consists of two stages. 
In a first stage, which takes place on a time scale of a few scattering times, 
the interaction energy roughly doubles. This can be understood by noticing that $E_\text{int}(t)$ depends on the fourth moment of the field, see Eq. (\ref{eq:interaction_energy}), which obeys: 
\begin{equation}\label{eq:psi4}
\average{\abs{\psi(x,t)}^4} = \average{\abs{\psi(x,t)}^2}^2 + \text{Var}\left(\abs{\psi(x,t)}^2\right).
\end{equation}
In the plane-wave limit where the profile is flat,
 the initial density variance is zero. 
 During the temporal evolution however, the density develops fluctuations depending on the realization of the disorder, which makes the interaction energy increase. 
The factor 2 enhancement is obtained by assuming
that, after a few scattering times, $\psi(x,t)$ is a complex Gaussian random variable.
The variance in Eq. (\ref{eq:psi4}) is then $\average{\abs{\psi(x,t)}^2}^2$, so that $\average{\abs{\psi(x,t)}^4}$ is doubled. This randomization of the complex wave function amplitude is similar to the appearance of optical speckles in scattering media~\cite{goodman2007speckle}. It implies that $E_\text{int}(t~\gg~\tau_0)\!=\!2 E_\text{int}(t~=~0)$ in the limit $\sigma=\infty$. For wave packets of finite size $\sigma,$ the effect is also present, albeit slightly smaller~
\footnote{The averaging must in principle be performed over many disorder realizations, while the interaction energy may be different for each disorder realization. In practice, a spatial averaging over the wave packet size is equivalent, provided the wave packet contains many speckle grains, which is the case if $\sigma\gg 1/k_0$.}. 

In the plane-wave limit $\sigma=\infty$, the interaction energy remains constant once the randomization process has ended. For finite $\sigma$ on the contrary, a second stage occurs, where $E_\text{int}$ decreases in time, see Fig. \ref{fig:speckelization}. This decrease is due to the spreading of the wave packet, which becomes more and more dilute. The spreading is initially fast, and then quickly slows down.

The time evolution of $E_\text{int}$ makes a detailed rescaling analysis of  the  boomerang effect very complex. The curves in Fig. \ref{fig:speckelization}, however, suggest the simple idea of using as scaling parameter the interaction energy \textit{at the break time $t_b$}, instead of the initial interaction energy. 
Although at finite $\sigma$ such a rescaling can only be approximate, since wave packets keep evolving slowly in time beyond $t_b$ (indicated  as arrows in Fig. \ref{fig:speckelization}), we show below that it provides satisfactory results.

Before applying this strategy, a final adjustment must be performed. The quantum boomerang being a \emph{dynamical} effect governed by the GPE~(\ref{eq:gpe}), its evolution is not strictly governed by  the interaction energy, $g\abs{\psi(x,t)}^4/2$, but rather by the \textit{nonlinear energy}, defined as
\begin{equation}\label{eq:nonlinear_energy}
E_\text{NL}(t) = 2E_\text{int}(t) = g\int\average{\abs{\psi(x,t)}^4}\diff{x}.
\end{equation}
In the appendix, we show that the $E_\text{NL}$ is related to the nonlinear self energy and discuss the dynamical behavior of the system at short time. We show analytically and numerically that it is indeed $E_\text{NL}$, rather than $E_\text{int}$, which governs the evolution. 

\subsection{Rescaling of the boomerang effect}

\begin{figure}
 \centering
 \includegraphics[width=\columnwidth]{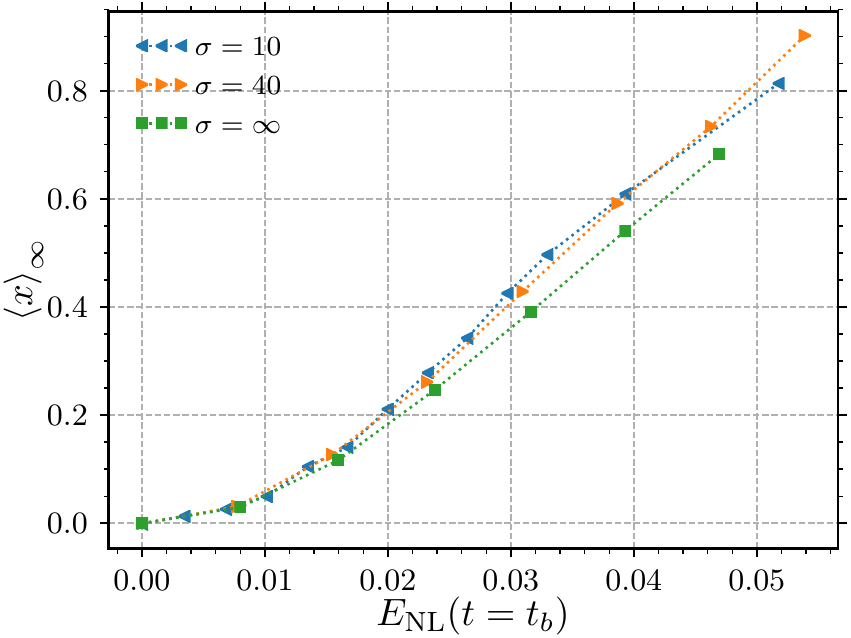}
 \caption{Long-time averages $\expval{x}_\infty$ of the CMP for different initial states vs. the nonlinear energy computed at the break time, $E_\text{NL}(t=t_b)$. $\expval{x}_\infty$ is expressed in units of $1/k_0$, while energy is in units of $\hbar^2k_0^2/m$. All results have been averaged over more than $5\times10^5$ disorder realizations. In this representation, data fall on the same single curve.}\label{fig:longtime_x_vs_nonlinear_energy}
\end{figure}

We can now re-analyze the boomerang effect using the nonlinear energy at the break time, $E_\text{NL}(t=t_b)$, as a control parameter of interactions.
We show  in Fig. \ref{fig:longtime_x_vs_nonlinear_energy} the long-time average $\expval{x}_\infty$ of the CMP as a function of the nonlinear energy calculated at the break time. In contrast with Fig. \ref{fig:longtime_x_vs_g_normalized}, now all points collapse on a single universal curve. As expected, in the regime of small nonlinear energy, $\expval{x}_\infty$ shows a quadratic dependence. 

\begin{figure}
 \centering
 \includegraphics[width=\columnwidth]{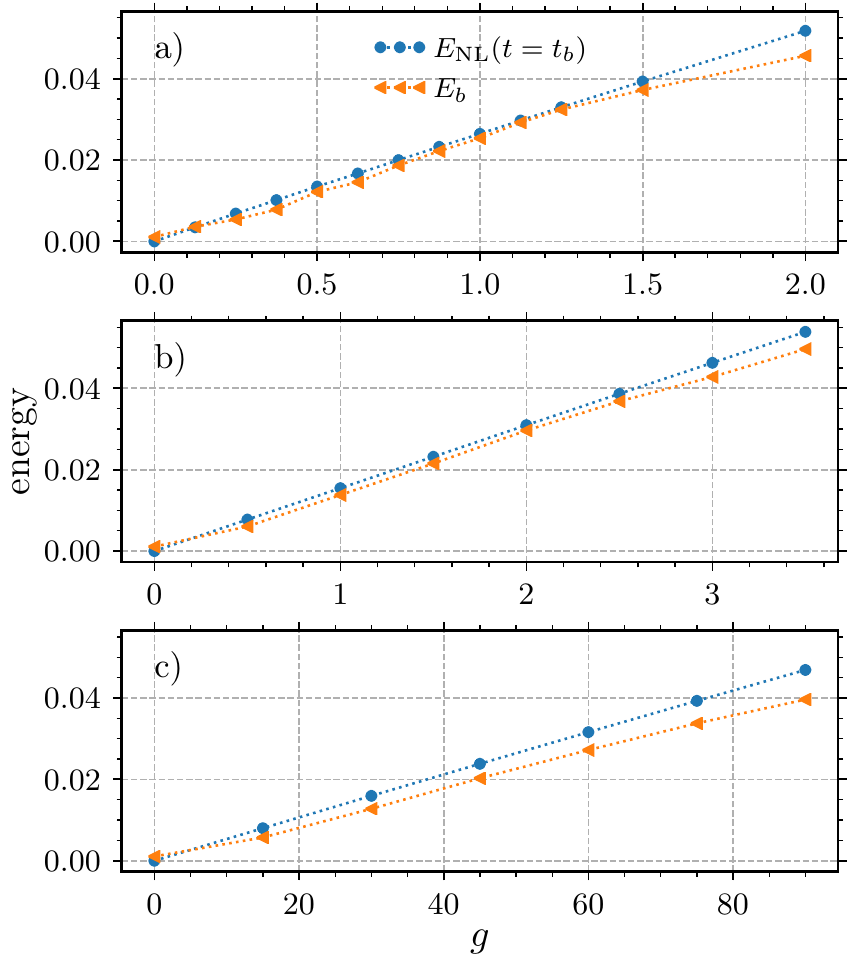}
 \caption{Comparison of the break energy, $E_b$, and the nonlinear energy at break time, $E_\text{NL}(t=t_b)$, for increasing values of the interaction strength. 
Results are shown for wave packets of size a) $\sigma=10/k_0$, b) $\sigma=40/k_0$ and c) $\sigma=\infty$.
Energy is in units of $\hbar^2k_0^2/m$, interaction strength $g$ in units of $\hbar^2 k_0/m$. All results have been averaged over $5\times10^5$ disorder realizations.}\label{fig:break_energy_vs_nonlinear_energy} 
\end{figure}

In Fig. \ref{fig:break_energy_vs_nonlinear_energy} we also compare the break energy $E_b=2\pi\hbar/t_b$ with the nonlinear energy at the break time $t_b$ 
for wave packets of size $\sigma=10/k_0,\,40/k_0$ and $\sigma=\infty$, for various values of $g$. There is a compelling evidence  that these quantities are very similar. A small difference is observed in the plane-wave limit, which we attribute to the slow residual decay of $\expval{x}$ due to early stage of thermalization, which leads to an underestimation of the break energy.
 Such a good agreement shows that a simple model involving a single parameter, the nonlinear energy at the break time, captures quantitatively the essential features of a complex dynamical process like the quantum boomerang effect for interacting particles. As discussed in the appendix, the same parameter turns out to also control the short-time dynamics and the correction of the scattering time due to interactions.

\section{Conclusion}
\label{sec:conclusion}
We have analyzed the quantum boomerang phenomenon in the presence of interactions on the basis of the one-dimensional Gross-Pitaevskii equation. 
We have found that weak  interactions do not destroy the quantum boomerang effect, in the sense that  the center of mass of a wave packet launched with a finite velocity is still retro-reflected after a few scattering times, slowly
returning towards its initial position. The boomerang effect is only partial though, as the quantum particle does not return to its initial position but stops on the way back. 
We have interpreted this phenomenon as a consequence of the destruction of interference between multiple scattering paths induced by a time-dependent nonlinear phase acquired along a path. To characterize this phenomenon, we have introduced a break time -- and the corresponding break energy -- the characteristic time beyond which the destruction of interference prevents
the wave packet to further move back to its initial position. 
We have finally shown that different initial states and interaction strengths can all be described by means of a single parameter, the nonlinear energy estimated at the break time.  

Our analysis is limited to the regime of weak disorder and weak interactions. When the disorder strength is increased, the quantitative description becomes more complicated, but the overall conclusions are expected to be qualitatively identical, provided interactions remain weak.  Indeed, a wave-packet in a relatively strong disorder contains many energy components, each energy being characterized by a scattering time and a mean free path. In the non-interacting limit, each energy component will display a boomerang effect described by Eq.~(\ref{eq:analytic_x}), but the superposition of various energy components will lead to a complicated $\expval{x(t)}$ function. In the presence of weak interactions, each energy component will display a partial boomerang effect, to that $\expval{x}_\infty$ is again likely to be nonzero. Whether an effective break time can be defined in such a case is an open question.
For weak disorder and stronger interactions, the break time is likely to decrease until it becomes comparable to the scattering time. Whether a single parameter also controls this regime is an interesting question left for future studies. 
Another important question -- especially if one envisions experiments with ultra-cold atoms -- is to know whether the observed softening of the boomerang effect due to interactions remains valid beyond the mean-field description. Studying the full many-body boomerang effect is a challenging task. 

\begin{acknowledgments}
We kindly thank PL-Grid Infrastructure for providing computational resources.
JJ and JZ acknowledge support of National Science Centre (Poland) under the project 
OPUS11 2016/21/B/ST2/01086. NC acknowledges financial support from the Agence Nationale de la Recherche (grant ANR-19-CE30-0028-01 CONFOCAL).  We also acknowledge support of Polish-French bilateral project Polonium 40490ZE. 
\end{acknowledgments}

\appendix
\section{Short time behavior}

We have seen that interactions modify the long-time dynamics of the center of mass position, and that a change of either parameters $\sigma$ or $g$ can be encompassed in the use of the nonlinear energy. In this appendix, we show that the very same concept -- the nonlinear time-dependent energy --  also accurately describes the short-time dynamics of the system, through a change of the real part of the self-energy and of the scattering time.

\subsection{Self energy in interacting systems}\label{sec:self_energy}

\begin{figure}
	\centering
	\includegraphics[width=\columnwidth]{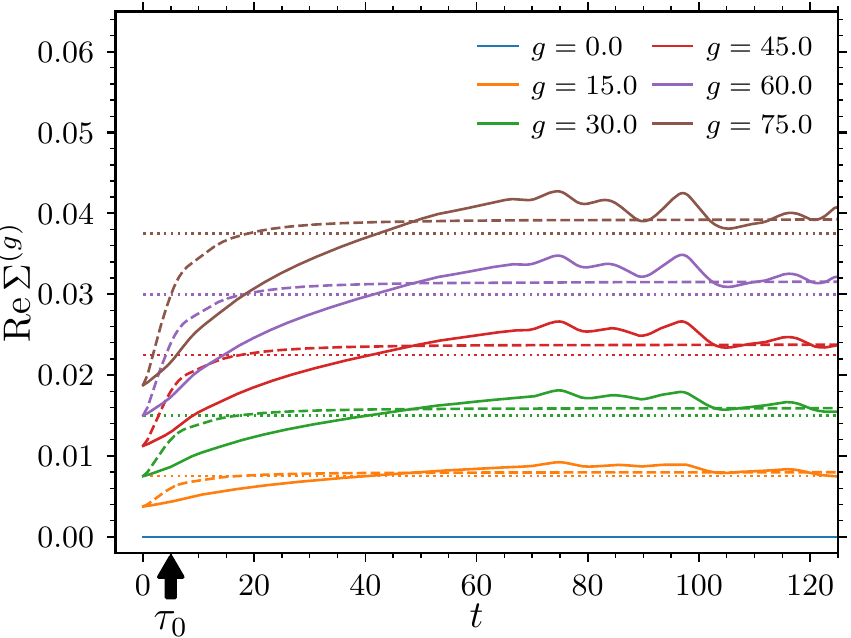}
	\caption{Numerically calculated real parts of the self energy $\Sigma^{(g)}$ (solid lines) for plane waves vs. time for several values of the interaction strength. In the plot we additionally show nonlinear energies $E_\text{NL}(t)$ (dashed lines), and indicate as dotted lines the value $2g\rho_0$. Energies are expressed in units of $\hbar^2k_0^2/m$ and time in units of $m/\hbar k_0^2$. The legend indicates curves from bottom to top. The value of the mean scattering time $\tau_0$ is indicated by the small black arrow near the time axis.} 
	\label{fig:self_energy}
\end{figure}

The self energy is a key concept in the description of disordered quantum systems. It is a complex valued function of the state energy $E$ and its momentum $\hbar k_0$ describing the disorder-induced energy shift and the exponential decay of average Green's functions in configuration space. It is noted $\Sigma_E(k_0)$ and defined by:
\be
\average{G}^{R}_E(k_0) = \frac{1}{E - E_0-\Sigma_E(k_0)},
\ee
where $\average{G}^{R}$ denotes the averaged retarded Green's function and $E_0=\hbar^2k_0^2/2m$ the disorder-free energy. The self-energy vanishes in the absence of disorder and is much smaller than $E_0$ in the weak-disorder limit (by a factor $1/k_0\ell_0$) \cite{akkermans2007mesoscopic}.

The evolution operator is the temporal Fourier transform of the Green's function. If the self-energy is a smooth function of $E$, one obtains for the evolution of a plane wave:
\begin{eqnarray}\label{eq:autocorrelation}
\average{\braket{\psi_0 | \psi(t)}} & = & \mathrm{e}^{-i\left(E_{k_0}+\Sigma_E(k_0)\right)t/\hbar}\nonumber\\ & = &\mathrm{e}^{-i\left(E_{k_0}+\Re\Sigma_{E_0}(k_0)\right)t/\hbar}\ \mathrm{e}^{\Im\Sigma_{E_0}t/\hbar},
\end{eqnarray}
so that $\Re\Sigma$ is an energy shift and $-\Im\Sigma$ the decay rate induced by the disorder. At the Born approximation, we have in 1D:
\be
\Sigma_{E_0}^{(0)}(k_0) = -\frac{i\hbar}{2\tau_0},
\ee
where the superscript $(0)$ refers to zero interactions.

In the presence of interactions, the situation is in general much more complicated.
Because of the nonlinearity of the GPE, the notion of evolution operator no longer exists
and the overlap Eq. (\ref{eq:autocorrelation}) has no reason to be an exponential function of time.
However, it is possible to define an effective self-energy using Eq. (\ref{eq:autocorrelation}), the left-hand-side of the equation being computed numerically from the solution $|\psi(t)\rangle$ of the GPE.
The obtained self-energy $\Sigma_{E_0}^{(g)}(k_0)$ depends on time.

 To analyze the impact of interactions on the self energy, we introduce its nonlinear part $\Sigma^{(g)}$, defined as:
\begin{equation}
\Sigma^{(g)} = \Sigma_{E_0}^{(g)}(k_0) - \Sigma_{E_0}^{(0)}(k_0),
\end{equation}
where both $\Sigma_{E_0}^{(g)}(k_0)$ and $\Sigma_{E_0}^{(0)}(k_0)$ are calculated numerically.
The real part of this quantity is plotted in Fig. \ref{fig:self_energy} in the plane-wave limit $\sigma=\infty$. 
 $\Sigma^{(g)}$ increases over a few mean scattering times and  then saturates at roughly twice its initial value.

It is easy to compute $\Sigma^{(g)}$ at $t=0$ from the GPE. The result is:
 \begin{equation}
 \Sigma^{(g)}(t=0) = E_\text{NL}(t=0) = g\rho_0,
 \label{eq:sigma0}
 \end{equation} 
where the first equality is valid for any initial state, while the second holds only for a plane wave.
At time longer than the scattering time, the randomization of the wave function phenomenon described in  Sec. \ref{subsec:nonlinear_energy} is responsible for a doubling of the nonlinear energy. It is thus very natural, and fully confirmed by the numerical results in~Fig. (\ref{fig:self_energy}) as well as by a  theoretical approach~\cite{wellens2019private} to have:
 \begin{equation}
\Sigma^{(g)}(t\gg \tau_0) = E_\text{NL}(t\gg \tau_0) = 2 g\rho_0.
 \label{eq:sigmainf}
\end{equation} 
This close connection between the nonlinear energy and the nonlinear part of the self-energy also exists at intermediate times, as shown in Fig. \ref{fig:self_energy}, where
we also plot numerically computed nonlinear energies. 
After an initial growth, both $\Sigma^{(g)}(t)$ and $E_\text{NL}(t)$ saturate around $2g\rho_0$ and follow a close evolution (even though the growth rate of $\Re\Sigma^{(g)}$ is slightly lower than for the nonlinear energy).
Altogether, it suggests that $\Re\Sigma^{(g)}$ and $E_\text{NL}$ may have a similar status for the problem of interacting disordered systems. This corroborates the conclusion of Sec. \ref{subsec:nonlinear_energy}, since $\Re\Sigma$ is typically  involved in the calculation of any observable, in particular of the CMP.

\subsection{Modification of the scattering time}
\label{subsec:short_time}

We now show that the nonlinear energy, Eq. (\ref{eq:nonlinear_energy}), which governs the long-time behavior of the quantum boomerang effect, also controls the change in the mean scattering time.

\begin{figure}
	\centering
	\includegraphics[width=\columnwidth]{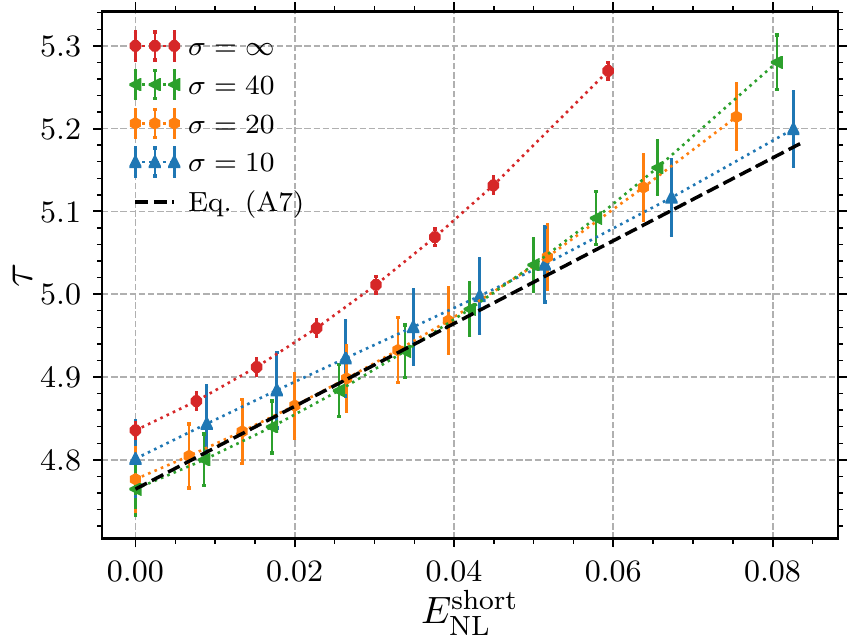}
	\caption{Fitted values of the mean scattering time $\tau$ for wave packets of different initial widths vs. the nonlinear energy averaged over the fit time window, $E_\text{NL}^\text{short}$. Time is in units of $m/\hbar k_0^2$ and energy in units of $\hbar^2 k_0^2/m$.  The black dashed line shows the prediction of Eq. (\ref{eq:tau_expansion}), where  the wave packet with $\sigma=40/k_0$ is used for $\tau_0$. }
	\label{fig:tau_vs_eint}
\end{figure}

As described in Sec. \ref{sec:lessening}, during the first part of the time evolution, precisely in the range $t<20-30\tau_0$, see Fig. \ref{fig:com_sigma_10_tmax1200},  the CMP is essentially not modified by interactions. The only difference between the interacting and non-interacting cases is that the mean scattering time and mean free path are increased. We have used the theoretical prediction for  $\expval{x(t)}$ \cite{prat2019quantumboomerang} in the non-interacting limit, which depends on $\tau$ and $\ell$, to fit the data in the interacting case in the short-time region, and thus access the dependence of $\tau$ and $\ell$ on the interaction strength $g$. Fits have been performed including weights inversely proportional to the square of the statistical errors in the time window $t\in[0, t_\text{fit}]$. Our choice is $t_\text{fit} =20\tau_0$. This value is chosen such that the fits enclose the whole ballistic motion and the beginning of the reflection.
The value of $t_\text{fit}$ slightly influences the fitted parameters, but the changes are smaller than the error originating from the fitting procedure. 

The fits return both $\tau$ and $\ell$ for a given interaction strength $g$. This allows us to calculate the average velocity $v=\ell/\tau$. From our data, we observe that the average velocity remains almost unaffected across all studied interaction values, although it is a little higher than the predicted value $\hbar k_0/m$ at the Born approximation. This apparent discrepancy is caused by higher order corrections to the Born approximation and is of the order of $1/k_0\ell_0\ll1$. We can thus restrict the analysis to the mean scattering time $\tau$ only.

The fitted values of $\tau$ are shown in Fig. \ref{fig:tau_vs_eint} as a function of the nonlinear energy averaged over the fit time window, $E_\text{NL}^\text{short}$. 
To explain these curves, we expand the mean scattering time $\tau(E\simeq E_0+E_\text{NL})$ to leading order in $E_\text{NL}\ll E_0$, using the Born approximation, Eq. (\ref{eq:born}). This yields:
\begin{equation}\label{eq:tau_expansion}
\tau \simeq \tau_0 + \frac{\hbar}{2k_0\gamma}E_\text{NL}.
\end{equation} 
The linear increase of $\tau$ is well visible in Fig. \ref{fig:tau_vs_eint}. The fact that curves at different $\sigma$ are slightly shifted upwards is due to the (small) dependence of the $g-$independent part, $\tau_0$, on $\sigma$, which shows up beyond the Born approximation. Eq. (\ref{eq:tau_expansion}) is shown in Fig. \ref{fig:tau_vs_eint}, where  for $\tau_0$ we use the numerical value of the scattering time for $\sigma=40/k_0$ and $g=0$. The agreement between Eq. (\ref{eq:tau_expansion}) and the data is very good.  
At larger values of $E_\text{NL}^\text{short},$ the curves start to deviate from the linear behavior, bending upwards. This effect is smaller for decreasing wave-packet widths $\sigma$. The observed change corresponds to a relatively strong boomerang breakdown, so that the use of the non-interacting theoretical prediction for the fit becomes less reliable.

\bibliography{boomerang_2019}

\providecommand{\noopsort}[1]{}\providecommand{\singleletter}[1]{#1}%
\begin{thebibliography}{51}%
\makeatletter
\providecommand \@ifxundefined [1]{%
 \@ifx{#1\undefined}
}%
\providecommand \@ifnum [1]{%
 \ifnum #1\expandafter \@firstoftwo
 \else \expandafter \@secondoftwo
 \fi
}%
\providecommand \@ifx [1]{%
 \ifx #1\expandafter \@firstoftwo
 \else \expandafter \@secondoftwo
 \fi
}%
\providecommand \natexlab [1]{#1}%
\providecommand \enquote  [1]{``#1''}%
\providecommand \bibnamefont  [1]{#1}%
\providecommand \bibfnamefont [1]{#1}%
\providecommand \citenamefont [1]{#1}%
\providecommand \href@noop [0]{\@secondoftwo}%
\providecommand \href [0]{\begingroup \@sanitize@url \@href}%
\providecommand \@href[1]{\@@startlink{#1}\@@href}%
\providecommand \@@href[1]{\endgroup#1\@@endlink}%
\providecommand \@sanitize@url [0]{\catcode `\\12\catcode `\$12\catcode
  `\&12\catcode `\#12\catcode `\^12\catcode `\_12\catcode `\%12\relax}%
\providecommand \@@startlink[1]{}%
\providecommand \@@endlink[0]{}%
\providecommand \url  [0]{\begingroup\@sanitize@url \@url }%
\providecommand \@url [1]{\endgroup\@href {#1}{\urlprefix }}%
\providecommand \urlprefix  [0]{URL }%
\providecommand \Eprint [0]{\href }%
\providecommand \doibase [0]{http://dx.doi.org/}%
\providecommand \selectlanguage [0]{\@gobble}%
\providecommand \bibinfo  [0]{\@secondoftwo}%
\providecommand \bibfield  [0]{\@secondoftwo}%
\providecommand \translation [1]{[#1]}%
\providecommand \BibitemOpen [0]{}%
\providecommand \bibitemStop [0]{}%
\providecommand \bibitemNoStop [0]{.\EOS\space}%
\providecommand \EOS [0]{\spacefactor3000\relax}%
\providecommand \BibitemShut  [1]{\csname bibitem#1\endcsname}%
\let\auto@bib@innerbib\@empty
\bibitem [{\citenamefont {Anderson}(1958)}]{anderson1958absence}%
  \BibitemOpen
  \bibfield  {author} {\bibinfo {author} {\bibfnamefont {P.~W.}\ \bibnamefont
  {Anderson}},\ }\bibfield  {title} {\enquote {\bibinfo {title} {Absence of
  diffusion in certain random lattices},}\ }\href {\doibase
  10.1103/PhysRev.109.1492} {\bibfield  {journal} {\bibinfo  {journal} {Phys.
  Rev.}\ }\textbf {\bibinfo {volume} {109}},\ \bibinfo {pages} {1492--1505}
  (\bibinfo {year} {1958})}\BibitemShut {NoStop}%
\bibitem [{\citenamefont {Evers}\ and\ \citenamefont {Mirlin}(2008)}]{Evers08}%
  \BibitemOpen
  \bibfield  {author} {\bibinfo {author} {\bibfnamefont {Ferdinand}\
  \bibnamefont {Evers}}\ and\ \bibinfo {author} {\bibfnamefont {Alexander~D.}\
  \bibnamefont {Mirlin}},\ }\bibfield  {title} {\enquote {\bibinfo {title}
  {{A}nderson transitions},}\ }\href {\doibase 10.1103/RevModPhys.80.1355}
  {\bibfield  {journal} {\bibinfo  {journal} {Rev. Mod. Phys.}\ }\textbf
  {\bibinfo {volume} {80}},\ \bibinfo {pages} {1355--1417} (\bibinfo {year}
  {2008})}\BibitemShut {NoStop}%
\bibitem [{\citenamefont {Chabanov}\ \emph {et~al.}(2000)\citenamefont
  {Chabanov}, \citenamefont {Stoytchev},\ and\ \citenamefont
  {Genack}}]{chabanov2000statistical}%
  \BibitemOpen
  \bibfield  {author} {\bibinfo {author} {\bibfnamefont {AA}~\bibnamefont
  {Chabanov}}, \bibinfo {author} {\bibfnamefont {M}~\bibnamefont {Stoytchev}},
  \ and\ \bibinfo {author} {\bibfnamefont {AZ}~\bibnamefont {Genack}},\
  }\bibfield  {title} {\enquote {\bibinfo {title} {Statistical signatures of
  photon localization},}\ }\href {\doibase 10.1038/35009055} {\bibfield
  {journal} {\bibinfo  {journal} {Nature}\ }\textbf {\bibinfo {volume} {404}},\
  \bibinfo {pages} {850} (\bibinfo {year} {2000})}\BibitemShut {NoStop}%
\bibitem [{\citenamefont {Schwartz}\ \emph {et~al.}(2007)\citenamefont
  {Schwartz}, \citenamefont {Bartal}, \citenamefont {Fishman},\ and\
  \citenamefont {Segev}}]{schwartz2007transport}%
  \BibitemOpen
  \bibfield  {author} {\bibinfo {author} {\bibfnamefont {Tal}\ \bibnamefont
  {Schwartz}}, \bibinfo {author} {\bibfnamefont {Guy}\ \bibnamefont {Bartal}},
  \bibinfo {author} {\bibfnamefont {Shmuel}\ \bibnamefont {Fishman}}, \ and\
  \bibinfo {author} {\bibfnamefont {Mordechai}\ \bibnamefont {Segev}},\
  }\bibfield  {title} {\enquote {\bibinfo {title} {Transport and {A}nderson
  localization in disordered two-dimensional photonic lattices},}\ }\href
  {\doibase 10.1038/nature05623} {\bibfield  {journal} {\bibinfo  {journal}
  {Nature}\ }\textbf {\bibinfo {volume} {446}},\ \bibinfo {pages} {52}
  (\bibinfo {year} {2007})}\BibitemShut {NoStop}%
\bibitem [{\citenamefont {Hu}\ \emph {et~al.}(2008)\citenamefont {Hu},
  \citenamefont {Strybulevych}, \citenamefont {Page}, \citenamefont
  {Skipetrov},\ and\ \citenamefont {van Tiggelen}}]{hu2008localization}%
  \BibitemOpen
  \bibfield  {author} {\bibinfo {author} {\bibfnamefont {Hefei}\ \bibnamefont
  {Hu}}, \bibinfo {author} {\bibfnamefont {A}~\bibnamefont {Strybulevych}},
  \bibinfo {author} {\bibfnamefont {JH}~\bibnamefont {Page}}, \bibinfo {author}
  {\bibfnamefont {Sergey~E}\ \bibnamefont {Skipetrov}}, \ and\ \bibinfo
  {author} {\bibfnamefont {Bart~A}\ \bibnamefont {van Tiggelen}},\ }\bibfield
  {title} {\enquote {\bibinfo {title} {Localization of ultrasound in a
  three-dimensional elastic network},}\ }\href {\doibase 10.1038/nphys1101}
  {\bibfield  {journal} {\bibinfo  {journal} {Nature Physics}\ }\textbf
  {\bibinfo {volume} {4}},\ \bibinfo {pages} {945} (\bibinfo {year}
  {2008})}\BibitemShut {NoStop}%
\bibitem [{\citenamefont {Billy}\ \emph {et~al.}(2008)\citenamefont {Billy},
  \citenamefont {Josse}, \citenamefont {Zuo}, \citenamefont {Bernard},
  \citenamefont {Hambrecht}, \citenamefont {Lugan}, \citenamefont
  {Cl{\'e}ment}, \citenamefont {Sanchez-Palencia}, \citenamefont {Bouyer},\
  and\ \citenamefont {Aspect}}]{billy2008direct}%
  \BibitemOpen
  \bibfield  {author} {\bibinfo {author} {\bibfnamefont {Juliette}\
  \bibnamefont {Billy}}, \bibinfo {author} {\bibfnamefont {Vincent}\
  \bibnamefont {Josse}}, \bibinfo {author} {\bibfnamefont {Zhanchun}\
  \bibnamefont {Zuo}}, \bibinfo {author} {\bibfnamefont {Alain}\ \bibnamefont
  {Bernard}}, \bibinfo {author} {\bibfnamefont {Ben}\ \bibnamefont
  {Hambrecht}}, \bibinfo {author} {\bibfnamefont {Pierre}\ \bibnamefont
  {Lugan}}, \bibinfo {author} {\bibfnamefont {David}\ \bibnamefont
  {Cl{\'e}ment}}, \bibinfo {author} {\bibfnamefont {Laurent}\ \bibnamefont
  {Sanchez-Palencia}}, \bibinfo {author} {\bibfnamefont {Philippe}\
  \bibnamefont {Bouyer}}, \ and\ \bibinfo {author} {\bibfnamefont {Alain}\
  \bibnamefont {Aspect}},\ }\bibfield  {title} {\enquote {\bibinfo {title}
  {Direct observation of {A}nderson localization of matter waves in a
  controlled disorder},}\ }\href {\doibase 10.1038/nature07000} {\bibfield
  {journal} {\bibinfo  {journal} {Nature}\ }\textbf {\bibinfo {volume} {453}},\
  \bibinfo {pages} {891} (\bibinfo {year} {2008})}\BibitemShut {NoStop}%
\bibitem [{\citenamefont {Jendrzejewski}\ \emph {et~al.}(2012)\citenamefont
  {Jendrzejewski}, \citenamefont {Bernard}, \citenamefont {Mueller},
  \citenamefont {Cheinet}, \citenamefont {Josse}, \citenamefont {Piraud},
  \citenamefont {Pezz{\'e}}, \citenamefont {Sanchez-Palencia}, \citenamefont
  {Aspect},\ and\ \citenamefont {Bouyer}}]{jendrzejewski2012three}%
  \BibitemOpen
  \bibfield  {author} {\bibinfo {author} {\bibfnamefont {Fred}\ \bibnamefont
  {Jendrzejewski}}, \bibinfo {author} {\bibfnamefont {Alain}\ \bibnamefont
  {Bernard}}, \bibinfo {author} {\bibfnamefont {Killian}\ \bibnamefont
  {Mueller}}, \bibinfo {author} {\bibfnamefont {Patrick}\ \bibnamefont
  {Cheinet}}, \bibinfo {author} {\bibfnamefont {Vincent}\ \bibnamefont
  {Josse}}, \bibinfo {author} {\bibfnamefont {Marie}\ \bibnamefont {Piraud}},
  \bibinfo {author} {\bibfnamefont {Luca}\ \bibnamefont {Pezz{\'e}}}, \bibinfo
  {author} {\bibfnamefont {Laurent}\ \bibnamefont {Sanchez-Palencia}}, \bibinfo
  {author} {\bibfnamefont {Alain}\ \bibnamefont {Aspect}}, \ and\ \bibinfo
  {author} {\bibfnamefont {Philippe}\ \bibnamefont {Bouyer}},\ }\bibfield
  {title} {\enquote {\bibinfo {title} {Three-dimensional localization of
  ultracold atoms in an optical disordered potential},}\ }\href {\doibase
  10.1038/nphys2256} {\bibfield  {journal} {\bibinfo  {journal} {Nature
  Physics}\ }\textbf {\bibinfo {volume} {8}},\ \bibinfo {pages} {398} (\bibinfo
  {year} {2012})}\BibitemShut {NoStop}%
\bibitem [{\citenamefont {Manai}\ \emph {et~al.}(2015)\citenamefont {Manai},
  \citenamefont {Cl\'ement}, \citenamefont {Chicireanu}, \citenamefont
  {Hainaut}, \citenamefont {Garreau}, \citenamefont {Szriftgiser},\ and\
  \citenamefont {Delande}}]{manai2015experimental}%
  \BibitemOpen
  \bibfield  {author} {\bibinfo {author} {\bibfnamefont {Isam}\ \bibnamefont
  {Manai}}, \bibinfo {author} {\bibfnamefont {Jean-Fran\c{c}ois}\ \bibnamefont
  {Cl\'ement}}, \bibinfo {author} {\bibfnamefont {Radu}\ \bibnamefont
  {Chicireanu}}, \bibinfo {author} {\bibfnamefont {Cl\'ement}\ \bibnamefont
  {Hainaut}}, \bibinfo {author} {\bibfnamefont {Jean~Claude}\ \bibnamefont
  {Garreau}}, \bibinfo {author} {\bibfnamefont {Pascal}\ \bibnamefont
  {Szriftgiser}}, \ and\ \bibinfo {author} {\bibfnamefont {Dominique}\
  \bibnamefont {Delande}},\ }\bibfield  {title} {\enquote {\bibinfo {title}
  {Experimental observation of two-dimensional {A}nderson localization with the
  atomic kicked rotor},}\ }\href {\doibase 10.1103/PhysRevLett.115.240603}
  {\bibfield  {journal} {\bibinfo  {journal} {Phys. Rev. Lett.}\ }\textbf
  {\bibinfo {volume} {115}},\ \bibinfo {pages} {240603} (\bibinfo {year}
  {2015})}\BibitemShut {NoStop}%
\bibitem [{\citenamefont {Semeghini}\ \emph {et~al.}(2015)\citenamefont
  {Semeghini}, \citenamefont {Landini}, \citenamefont {Castilho}, \citenamefont
  {Roy}, \citenamefont {Spagnolli}, \citenamefont {Trenkwalder}, \citenamefont
  {Fattori}, \citenamefont {Inguscio},\ and\ \citenamefont
  {Modugno}}]{semeghini2015measurement}%
  \BibitemOpen
  \bibfield  {author} {\bibinfo {author} {\bibfnamefont {Giulia}\ \bibnamefont
  {Semeghini}}, \bibinfo {author} {\bibfnamefont {Manuele}\ \bibnamefont
  {Landini}}, \bibinfo {author} {\bibfnamefont {Patricia}\ \bibnamefont
  {Castilho}}, \bibinfo {author} {\bibfnamefont {Sanjukta}\ \bibnamefont
  {Roy}}, \bibinfo {author} {\bibfnamefont {Giacomo}\ \bibnamefont
  {Spagnolli}}, \bibinfo {author} {\bibfnamefont {Andreas}\ \bibnamefont
  {Trenkwalder}}, \bibinfo {author} {\bibfnamefont {Marco}\ \bibnamefont
  {Fattori}}, \bibinfo {author} {\bibfnamefont {Massimo}\ \bibnamefont
  {Inguscio}}, \ and\ \bibinfo {author} {\bibfnamefont {Giovanni}\ \bibnamefont
  {Modugno}},\ }\bibfield  {title} {\enquote {\bibinfo {title} {Measurement of
  the mobility edge for 3d {A}nderson localization},}\ }\href {\doibase
  10.1038/nphys3339} {\bibfield  {journal} {\bibinfo  {journal} {Nature
  Physics}\ }\textbf {\bibinfo {volume} {11}},\ \bibinfo {pages} {554}
  (\bibinfo {year} {2015})}\BibitemShut {NoStop}%
\bibitem [{\citenamefont {White}\ \emph {et~al.}(2019)\citenamefont {White},
  \citenamefont {Haase}, \citenamefont {Brown}, \citenamefont {Hoogerland},
  \citenamefont {Najafabadi}, \citenamefont {Helm}, \citenamefont {Gies},
  \citenamefont {Schumayer},\ and\ \citenamefont
  {Hutchinson}}]{white2019observation}%
  \BibitemOpen
  \bibfield  {author} {\bibinfo {author} {\bibfnamefont {Donald~H}\
  \bibnamefont {White}}, \bibinfo {author} {\bibfnamefont {Thomas~A}\
  \bibnamefont {Haase}}, \bibinfo {author} {\bibfnamefont {Dylan~J}\
  \bibnamefont {Brown}}, \bibinfo {author} {\bibfnamefont {Maarten~D}\
  \bibnamefont {Hoogerland}}, \bibinfo {author} {\bibfnamefont {Mojdeh~S}\
  \bibnamefont {Najafabadi}}, \bibinfo {author} {\bibfnamefont {John~L}\
  \bibnamefont {Helm}}, \bibinfo {author} {\bibfnamefont {Christopher}\
  \bibnamefont {Gies}}, \bibinfo {author} {\bibfnamefont {Daniel}\ \bibnamefont
  {Schumayer}}, \ and\ \bibinfo {author} {\bibfnamefont {David~AW}\
  \bibnamefont {Hutchinson}},\ }\bibfield  {title} {\enquote {\bibinfo {title}
  {Observation of two-dimensional {A}nderson localisation of ultracold
  atoms},}\ }\href {https://arxiv.org/abs/1911.04858} {\bibfield  {journal}
  {\bibinfo  {journal} {arXiv preprint arXiv:1911.04858}\ } (\bibinfo {year}
  {2019})}\BibitemShut {NoStop}%
\bibitem [{\citenamefont {Aubry}\ and\ \citenamefont
  {Andr{\'e}}(1980)}]{Aubry80}%
  \BibitemOpen
  \bibfield  {author} {\bibinfo {author} {\bibfnamefont {Serge}\ \bibnamefont
  {Aubry}}\ and\ \bibinfo {author} {\bibfnamefont {Gilles}\ \bibnamefont
  {Andr{\'e}}},\ }\bibfield  {title} {\enquote {\bibinfo {title} {Analyticity
  breaking and {A}nderson localization in incommensurate lattices},}\
  }\href@noop {} {\bibfield  {journal} {\bibinfo  {journal} {Ann. Israel Phys.
  Soc}\ }\textbf {\bibinfo {volume} {3}},\ \bibinfo {pages} {18} (\bibinfo
  {year} {1980})}\BibitemShut {NoStop}%
\bibitem [{\citenamefont {Roati}\ \emph {et~al.}(2008)\citenamefont {Roati},
  \citenamefont {D’Errico}, \citenamefont {Fallani}, \citenamefont {Fattori},
  \citenamefont {Fort}, \citenamefont {Zaccanti}, \citenamefont {Modugno},\
  and\ \citenamefont {Inguscio}}]{Roati08}%
  \BibitemOpen
  \bibfield  {author} {\bibinfo {author} {\bibfnamefont {Giacomo}\ \bibnamefont
  {Roati}}, \bibinfo {author} {\bibfnamefont {Chiara}\ \bibnamefont
  {D’Errico}}, \bibinfo {author} {\bibfnamefont {Leonardo}\ \bibnamefont
  {Fallani}}, \bibinfo {author} {\bibfnamefont {Marco}\ \bibnamefont
  {Fattori}}, \bibinfo {author} {\bibfnamefont {Chiara}\ \bibnamefont {Fort}},
  \bibinfo {author} {\bibfnamefont {Matteo}\ \bibnamefont {Zaccanti}}, \bibinfo
  {author} {\bibfnamefont {Giovanni}\ \bibnamefont {Modugno}}, \ and\ \bibinfo
  {author} {\bibfnamefont {Massimo}\ \bibnamefont {Inguscio}},\ }\bibfield
  {title} {\enquote {\bibinfo {title} {{A}nderson localization of a
  non-interacting {B}ose–{E}instein condensate},}\ }\href {\doibase
  10.1038/nature07071} {\bibfield  {journal} {\bibinfo  {journal} {Nature}\
  }\textbf {\bibinfo {volume} {453}},\ \bibinfo {pages} {895} (\bibinfo {year}
  {2008})}\BibitemShut {NoStop}%
\bibitem [{\citenamefont {Gornyi}\ \emph {et~al.}(2005)\citenamefont {Gornyi},
  \citenamefont {Mirlin},\ and\ \citenamefont {Polyakov}}]{Gornyi05}%
  \BibitemOpen
  \bibfield  {author} {\bibinfo {author} {\bibfnamefont {I.~V.}\ \bibnamefont
  {Gornyi}}, \bibinfo {author} {\bibfnamefont {A.~D.}\ \bibnamefont {Mirlin}},
  \ and\ \bibinfo {author} {\bibfnamefont {D.~G.}\ \bibnamefont {Polyakov}},\
  }\bibfield  {title} {\enquote {\bibinfo {title} {Interacting electrons in
  disordered wires: {A}nderson localization and low-$t$ transport},}\ }\href
  {\doibase 10.1103/PhysRevLett.95.206603} {\bibfield  {journal} {\bibinfo
  {journal} {Phys. Rev. Lett.}\ }\textbf {\bibinfo {volume} {95}},\ \bibinfo
  {pages} {206603} (\bibinfo {year} {2005})}\BibitemShut {NoStop}%
\bibitem [{\citenamefont {Basko}\ \emph {et~al.}(2006)\citenamefont {Basko},
  \citenamefont {Aleiner},\ and\ \citenamefont {Altshuler}}]{basko2006metal}%
  \BibitemOpen
  \bibfield  {author} {\bibinfo {author} {\bibfnamefont {Denis~M}\ \bibnamefont
  {Basko}}, \bibinfo {author} {\bibfnamefont {Igor~L}\ \bibnamefont {Aleiner}},
  \ and\ \bibinfo {author} {\bibfnamefont {Boris~L}\ \bibnamefont
  {Altshuler}},\ }\bibfield  {title} {\enquote {\bibinfo {title}
  {Metal--insulator transition in a weakly interacting many-electron system
  with localized single-particle states},}\ }\href {\doibase
  10.1016/j.aop.2005.11.014} {\bibfield  {journal} {\bibinfo  {journal} {Ann.
  Phys. (N.Y.)}\ }\textbf {\bibinfo {volume} {321}},\ \bibinfo {pages}
  {1126--1205} (\bibinfo {year} {2006})}\BibitemShut {NoStop}%
\bibitem [{\citenamefont {Huse}\ \emph {et~al.}(2014)\citenamefont {Huse},
  \citenamefont {Nandkishore},\ and\ \citenamefont {Oganesyan}}]{Huse14}%
  \BibitemOpen
  \bibfield  {author} {\bibinfo {author} {\bibfnamefont {David~A.}\
  \bibnamefont {Huse}}, \bibinfo {author} {\bibfnamefont {Rahul}\ \bibnamefont
  {Nandkishore}}, \ and\ \bibinfo {author} {\bibfnamefont {Vadim}\ \bibnamefont
  {Oganesyan}},\ }\bibfield  {title} {\enquote {\bibinfo {title} {Phenomenology
  of fully many-body-localized systems},}\ }\href {\doibase
  10.1103/PhysRevB.90.174202} {\bibfield  {journal} {\bibinfo  {journal} {Phys.
  Rev. B}\ }\textbf {\bibinfo {volume} {90}},\ \bibinfo {pages} {174202}
  (\bibinfo {year} {2014})}\BibitemShut {NoStop}%
\bibitem [{\citenamefont {Alet}\ and\ \citenamefont
  {Laflorencie}(2018)}]{Alet18}%
  \BibitemOpen
  \bibfield  {author} {\bibinfo {author} {\bibfnamefont {Fabien}\ \bibnamefont
  {Alet}}\ and\ \bibinfo {author} {\bibfnamefont {Nicolas}\ \bibnamefont
  {Laflorencie}},\ }\bibfield  {title} {\enquote {\bibinfo {title} {Many-body
  localization: An introduction and selected topics},}\ }\href {\doibase
  10.1016/j.crhy.2018.03.003} {\bibfield  {journal} {\bibinfo  {journal}
  {Comptes Rendus Physique}\ }\textbf {\bibinfo {volume} {19}},\ \bibinfo
  {pages} {498 -- 525} (\bibinfo {year} {2018})},\ \bibinfo {note} {{Q}uantum
  simulation / {S}imulation quantique}\BibitemShut {NoStop}%
\bibitem [{\citenamefont {Schreiber}\ \emph {et~al.}(2015)\citenamefont
  {Schreiber}, \citenamefont {Hodgman}, \citenamefont {Bordia}, \citenamefont
  {L{\"u}schen}, \citenamefont {Fischer}, \citenamefont {Vosk}, \citenamefont
  {Altman}, \citenamefont {Schneider},\ and\ \citenamefont
  {Bloch}}]{schreiber2015observation}%
  \BibitemOpen
  \bibfield  {author} {\bibinfo {author} {\bibfnamefont {Michael}\ \bibnamefont
  {Schreiber}}, \bibinfo {author} {\bibfnamefont {Sean~S.}\ \bibnamefont
  {Hodgman}}, \bibinfo {author} {\bibfnamefont {Pranjal}\ \bibnamefont
  {Bordia}}, \bibinfo {author} {\bibfnamefont {Henrik~P.}\ \bibnamefont
  {L{\"u}schen}}, \bibinfo {author} {\bibfnamefont {Mark~H.}\ \bibnamefont
  {Fischer}}, \bibinfo {author} {\bibfnamefont {Ronen}\ \bibnamefont {Vosk}},
  \bibinfo {author} {\bibfnamefont {Ehud}\ \bibnamefont {Altman}}, \bibinfo
  {author} {\bibfnamefont {Ulrich}\ \bibnamefont {Schneider}}, \ and\ \bibinfo
  {author} {\bibfnamefont {Immanuel}\ \bibnamefont {Bloch}},\ }\bibfield
  {title} {\enquote {\bibinfo {title} {Observation of many-body localization of
  interacting fermions in a quasirandom optical lattice},}\ }\href {\doibase
  10.1126/science.aaa7432} {\bibfield  {journal} {\bibinfo  {journal}
  {Science}\ }\textbf {\bibinfo {volume} {349}},\ \bibinfo {pages} {842--845}
  (\bibinfo {year} {2015})}\BibitemShut {NoStop}%
\bibitem [{\citenamefont {Choi}\ \emph {et~al.}(2016)\citenamefont {Choi},
  \citenamefont {Hild}, \citenamefont {Zeiher}, \citenamefont {Schau{\ss}},
  \citenamefont {Rubio-Abadal}, \citenamefont {Yefsah}, \citenamefont
  {Khemani}, \citenamefont {Huse}, \citenamefont {Bloch},\ and\ \citenamefont
  {Gross}}]{choi2016exploring}%
  \BibitemOpen
  \bibfield  {author} {\bibinfo {author} {\bibfnamefont {Jae-Yoon}\
  \bibnamefont {Choi}}, \bibinfo {author} {\bibfnamefont {Sebastian}\
  \bibnamefont {Hild}}, \bibinfo {author} {\bibfnamefont {Johannes}\
  \bibnamefont {Zeiher}}, \bibinfo {author} {\bibfnamefont {Peter}\
  \bibnamefont {Schau{\ss}}}, \bibinfo {author} {\bibfnamefont {Antonio}\
  \bibnamefont {Rubio-Abadal}}, \bibinfo {author} {\bibfnamefont {Tarik}\
  \bibnamefont {Yefsah}}, \bibinfo {author} {\bibfnamefont {Vedika}\
  \bibnamefont {Khemani}}, \bibinfo {author} {\bibfnamefont {David~A}\
  \bibnamefont {Huse}}, \bibinfo {author} {\bibfnamefont {Immanuel}\
  \bibnamefont {Bloch}}, \ and\ \bibinfo {author} {\bibfnamefont {Christian}\
  \bibnamefont {Gross}},\ }\bibfield  {title} {\enquote {\bibinfo {title}
  {Exploring the many-body localization transition in two dimensions},}\ }\href
  {\doibase 10.1126/science.aaf8834} {\bibfield  {journal} {\bibinfo  {journal}
  {Science}\ }\textbf {\bibinfo {volume} {352}},\ \bibinfo {pages} {1547--1552}
  (\bibinfo {year} {2016})}\BibitemShut {NoStop}%
\bibitem [{\citenamefont {{{\v{S}}untajs}}\ \emph {et~al.}(2019)\citenamefont
  {{{\v{S}}untajs}}, \citenamefont {{Bon{\v{c}}a}}, \citenamefont {{Prosen}},\
  and\ \citenamefont {{Vidmar}}}]{Suntajs19}%
  \BibitemOpen
  \bibfield  {author} {\bibinfo {author} {\bibfnamefont {J.}~\bibnamefont
  {{{\v{S}}untajs}}}, \bibinfo {author} {\bibfnamefont {J.}~\bibnamefont
  {{Bon{\v{c}}a}}}, \bibinfo {author} {\bibfnamefont {T.}~\bibnamefont
  {{Prosen}}}, \ and\ \bibinfo {author} {\bibfnamefont {L.}~\bibnamefont
  {{Vidmar}}},\ }\bibfield  {title} {\enquote {\bibinfo {title} {Quantum chaos
  challenges many-body localization},}\ }\href
  {https://arxiv.org/abs/1905.06345} {\bibfield  {journal} {\bibinfo  {journal}
  {arXiv preprint}\ ,\ \bibinfo {pages} {arXiv:1905.06345}} (\bibinfo {year}
  {2019})}\BibitemShut {NoStop}%
\bibitem [{\citenamefont {Abanin}\ \emph {et~al.}(2019)\citenamefont {Abanin},
  \citenamefont {Bardarson}, \citenamefont {Tomasi}, \citenamefont
  {Gopalakrishnan}, \citenamefont {Khemani}, \citenamefont {Parameswaran},
  \citenamefont {Pollmann}, \citenamefont {Potter}, \citenamefont {Serbyn},\
  and\ \citenamefont {Vasseur}}]{Abanin19z}%
  \BibitemOpen
  \bibfield  {author} {\bibinfo {author} {\bibfnamefont {D.~A.}\ \bibnamefont
  {Abanin}}, \bibinfo {author} {\bibfnamefont {J.~H.}\ \bibnamefont
  {Bardarson}}, \bibinfo {author} {\bibfnamefont {G.~De}\ \bibnamefont
  {Tomasi}}, \bibinfo {author} {\bibfnamefont {S.}~\bibnamefont
  {Gopalakrishnan}}, \bibinfo {author} {\bibfnamefont {V.}~\bibnamefont
  {Khemani}}, \bibinfo {author} {\bibfnamefont {S.~A.}\ \bibnamefont
  {Parameswaran}}, \bibinfo {author} {\bibfnamefont {F.}~\bibnamefont
  {Pollmann}}, \bibinfo {author} {\bibfnamefont {A.~C.}\ \bibnamefont
  {Potter}}, \bibinfo {author} {\bibfnamefont {M.}~\bibnamefont {Serbyn}}, \
  and\ \bibinfo {author} {\bibfnamefont {R.}~\bibnamefont {Vasseur}},\
  }\bibfield  {title} {\enquote {\bibinfo {title} {Distinguishing localization
  from chaos: challenges in finite-size systems},}\ }\href
  {https://arxiv.org/abs/1911.04501} {\bibfield  {journal} {\bibinfo  {journal}
  {arXiv preprint}\ ,\ \bibinfo {pages} {arXiv:1911.04501}} (\bibinfo {year}
  {2019})}\BibitemShut {NoStop}%
\bibitem [{\citenamefont {Sierant}\ \emph {et~al.}(2020)\citenamefont
  {Sierant}, \citenamefont {Delande},\ and\ \citenamefont
  {Zakrzewski}}]{Sierant19z}%
  \BibitemOpen
  \bibfield  {author} {\bibinfo {author} {\bibfnamefont {Piotr}\ \bibnamefont
  {Sierant}}, \bibinfo {author} {\bibfnamefont {Dominique}\ \bibnamefont
  {Delande}}, \ and\ \bibinfo {author} {\bibfnamefont {Jakub}\ \bibnamefont
  {Zakrzewski}},\ }\bibfield  {title} {\enquote {\bibinfo {title} {Thouless
  time analysis of {A}nderson and many-body localization transitions},}\ }\href
  {\doibase 10.1103/PhysRevLett.124.186601} {\bibfield  {journal} {\bibinfo
  {journal} {Phys. Rev. Lett.}\ }\textbf {\bibinfo {volume} {124}},\ \bibinfo
  {pages} {186601} (\bibinfo {year} {2020})}\BibitemShut {NoStop}%
\bibitem [{\citenamefont {Panda}\ \emph {et~al.}(2020)\citenamefont {Panda},
  \citenamefont {Scardicchio}, \citenamefont {Schulz}, \citenamefont {Taylor},\
  and\ \citenamefont {{\v{Z}}nidari{\v{c}}}}]{Panda19}%
  \BibitemOpen
  \bibfield  {author} {\bibinfo {author} {\bibfnamefont {R.~K.}\ \bibnamefont
  {Panda}}, \bibinfo {author} {\bibfnamefont {A.}~\bibnamefont {Scardicchio}},
  \bibinfo {author} {\bibfnamefont {M.}~\bibnamefont {Schulz}}, \bibinfo
  {author} {\bibfnamefont {S.~R.}\ \bibnamefont {Taylor}}, \ and\ \bibinfo
  {author} {\bibfnamefont {M.}~\bibnamefont {{\v{Z}}nidari{\v{c}}}},\
  }\bibfield  {title} {\enquote {\bibinfo {title} {Can we study the many-body
  localisation transition?}}\ }\href {\doibase 10.1209/0295-5075/128/67003}
  {\bibfield  {journal} {\bibinfo  {journal} {{EPL} (Europhysics Letters)}\
  }\textbf {\bibinfo {volume} {128}},\ \bibinfo {pages} {67003} (\bibinfo
  {year} {2020})}\BibitemShut {NoStop}%
\bibitem [{\citenamefont {L\"uschen}\ \emph {et~al.}(2018)\citenamefont
  {L\"uschen}, \citenamefont {Scherg}, \citenamefont {Kohlert}, \citenamefont
  {Schreiber}, \citenamefont {Bordia}, \citenamefont {Li}, \citenamefont
  {Das~Sarma},\ and\ \citenamefont {Bloch}}]{Luschen18}%
  \BibitemOpen
  \bibfield  {author} {\bibinfo {author} {\bibfnamefont {Henrik~P.}\
  \bibnamefont {L\"uschen}}, \bibinfo {author} {\bibfnamefont {Sebastian}\
  \bibnamefont {Scherg}}, \bibinfo {author} {\bibfnamefont {Thomas}\
  \bibnamefont {Kohlert}}, \bibinfo {author} {\bibfnamefont {Michael}\
  \bibnamefont {Schreiber}}, \bibinfo {author} {\bibfnamefont {Pranjal}\
  \bibnamefont {Bordia}}, \bibinfo {author} {\bibfnamefont {Xiao}\ \bibnamefont
  {Li}}, \bibinfo {author} {\bibfnamefont {S.}~\bibnamefont {Das~Sarma}}, \
  and\ \bibinfo {author} {\bibfnamefont {Immanuel}\ \bibnamefont {Bloch}},\
  }\bibfield  {title} {\enquote {\bibinfo {title} {Single-particle mobility
  edge in a one-dimensional quasiperiodic optical lattice},}\ }\href {\doibase
  10.1103/PhysRevLett.120.160404} {\bibfield  {journal} {\bibinfo  {journal}
  {Phys. Rev. Lett.}\ }\textbf {\bibinfo {volume} {120}},\ \bibinfo {pages}
  {160404} (\bibinfo {year} {2018})}\BibitemShut {NoStop}%
\bibitem [{\citenamefont {Major}\ \emph {et~al.}(2018)\citenamefont {Major},
  \citenamefont {Morigi},\ and\ \citenamefont {Zakrzewski}}]{Major18}%
  \BibitemOpen
  \bibfield  {author} {\bibinfo {author} {\bibfnamefont {Jan}\ \bibnamefont
  {Major}}, \bibinfo {author} {\bibfnamefont {Giovanna}\ \bibnamefont
  {Morigi}}, \ and\ \bibinfo {author} {\bibfnamefont {Jakub}\ \bibnamefont
  {Zakrzewski}},\ }\bibfield  {title} {\enquote {\bibinfo {title}
  {Single-particle localization in dynamical potentials},}\ }\href {\doibase
  10.1103/PhysRevA.98.053633} {\bibfield  {journal} {\bibinfo  {journal} {Phys.
  Rev. A}\ }\textbf {\bibinfo {volume} {98}},\ \bibinfo {pages} {053633}
  (\bibinfo {year} {2018})}\BibitemShut {NoStop}%
\bibitem [{\citenamefont {Prat}\ \emph {et~al.}(2019)\citenamefont {Prat},
  \citenamefont {Delande},\ and\ \citenamefont
  {Cherroret}}]{prat2019quantumboomerang}%
  \BibitemOpen
  \bibfield  {author} {\bibinfo {author} {\bibfnamefont {Tony}\ \bibnamefont
  {Prat}}, \bibinfo {author} {\bibfnamefont {Dominique}\ \bibnamefont
  {Delande}}, \ and\ \bibinfo {author} {\bibfnamefont {Nicolas}\ \bibnamefont
  {Cherroret}},\ }\bibfield  {title} {\enquote {\bibinfo {title} {Quantum
  boomeranglike effect of wave packets in random media},}\ }\href {\doibase
  10.1103/PhysRevA.99.023629} {\bibfield  {journal} {\bibinfo  {journal} {Phys.
  Rev. A}\ }\textbf {\bibinfo {volume} {99}},\ \bibinfo {pages} {023629}
  (\bibinfo {year} {2019})}\BibitemShut {NoStop}%
\bibitem [{\citenamefont {Pikovsky}\ and\ \citenamefont
  {Shepelyansky}(2008)}]{pikovsky2008destruction}%
  \BibitemOpen
  \bibfield  {author} {\bibinfo {author} {\bibfnamefont {A.~S.}\ \bibnamefont
  {Pikovsky}}\ and\ \bibinfo {author} {\bibfnamefont {D.~L.}\ \bibnamefont
  {Shepelyansky}},\ }\bibfield  {title} {\enquote {\bibinfo {title}
  {Destruction of {A}nderson localization by a weak nonlinearity},}\ }\href
  {\doibase 10.1103/PhysRevLett.100.094101} {\bibfield  {journal} {\bibinfo
  {journal} {Phys. Rev. Lett.}\ }\textbf {\bibinfo {volume} {100}},\ \bibinfo
  {pages} {094101} (\bibinfo {year} {2008})}\BibitemShut {NoStop}%
\bibitem [{\citenamefont {Skokos}\ \emph {et~al.}(2009)\citenamefont {Skokos},
  \citenamefont {Krimer}, \citenamefont {Komineas},\ and\ \citenamefont
  {Flach}}]{skokos2009delocalization}%
  \BibitemOpen
  \bibfield  {author} {\bibinfo {author} {\bibfnamefont {Ch.}\ \bibnamefont
  {Skokos}}, \bibinfo {author} {\bibfnamefont {D.~O.}\ \bibnamefont {Krimer}},
  \bibinfo {author} {\bibfnamefont {S.}~\bibnamefont {Komineas}}, \ and\
  \bibinfo {author} {\bibfnamefont {S.}~\bibnamefont {Flach}},\ }\bibfield
  {title} {\enquote {\bibinfo {title} {Delocalization of wave packets in
  disordered nonlinear chains},}\ }\href {\doibase 10.1103/PhysRevE.79.056211}
  {\bibfield  {journal} {\bibinfo  {journal} {Phys. Rev. E}\ }\textbf {\bibinfo
  {volume} {79}},\ \bibinfo {pages} {056211} (\bibinfo {year}
  {2009})}\BibitemShut {NoStop}%
\bibitem [{\citenamefont {Flach}\ \emph {et~al.}(2009)\citenamefont {Flach},
  \citenamefont {Krimer},\ and\ \citenamefont {Skokos}}]{flach2009universal}%
  \BibitemOpen
  \bibfield  {author} {\bibinfo {author} {\bibfnamefont {S.}~\bibnamefont
  {Flach}}, \bibinfo {author} {\bibfnamefont {D.~O.}\ \bibnamefont {Krimer}}, \
  and\ \bibinfo {author} {\bibfnamefont {Ch.}\ \bibnamefont {Skokos}},\
  }\bibfield  {title} {\enquote {\bibinfo {title} {Universal spreading of wave
  packets in disordered nonlinear systems},}\ }\href {\doibase
  10.1103/PhysRevLett.102.024101} {\bibfield  {journal} {\bibinfo  {journal}
  {Phys. Rev. Lett.}\ }\textbf {\bibinfo {volume} {102}},\ \bibinfo {pages}
  {024101} (\bibinfo {year} {2009})}\BibitemShut {NoStop}%
\bibitem [{\citenamefont {Cherroret}\ \emph {et~al.}(2014)\citenamefont
  {Cherroret}, \citenamefont {Vermersch}, \citenamefont {Garreau},\ and\
  \citenamefont {Delande}}]{Cherroret14}%
  \BibitemOpen
  \bibfield  {author} {\bibinfo {author} {\bibfnamefont {Nicolas}\ \bibnamefont
  {Cherroret}}, \bibinfo {author} {\bibfnamefont {Beno\^{\i}t}\ \bibnamefont
  {Vermersch}}, \bibinfo {author} {\bibfnamefont {Jean~Claude}\ \bibnamefont
  {Garreau}}, \ and\ \bibinfo {author} {\bibfnamefont {Dominique}\ \bibnamefont
  {Delande}},\ }\bibfield  {title} {\enquote {\bibinfo {title} {How nonlinear
  interactions challenge the three-dimensional {A}nderson transition},}\ }\href
  {\doibase 10.1103/PhysRevLett.112.170603} {\bibfield  {journal} {\bibinfo
  {journal} {Phys. Rev. Lett.}\ }\textbf {\bibinfo {volume} {112}},\ \bibinfo
  {pages} {170603} (\bibinfo {year} {2014})}\BibitemShut {NoStop}%
\bibitem [{\citenamefont {Vakulchyk}\ \emph {et~al.}(2019)\citenamefont
  {Vakulchyk}, \citenamefont {Fistul},\ and\ \citenamefont
  {Flach}}]{vakulchyk2018universal}%
  \BibitemOpen
  \bibfield  {author} {\bibinfo {author} {\bibfnamefont {Ihor}\ \bibnamefont
  {Vakulchyk}}, \bibinfo {author} {\bibfnamefont {Mikhail~V.}\ \bibnamefont
  {Fistul}}, \ and\ \bibinfo {author} {\bibfnamefont {Sergej}\ \bibnamefont
  {Flach}},\ }\bibfield  {title} {\enquote {\bibinfo {title} {Wave packet
  spreading with disordered nonlinear discrete-time quantum walks},}\ }\href
  {\doibase 10.1103/PhysRevLett.122.040501} {\bibfield  {journal} {\bibinfo
  {journal} {Phys. Rev. Lett.}\ }\textbf {\bibinfo {volume} {122}},\ \bibinfo
  {pages} {040501} (\bibinfo {year} {2019})}\BibitemShut {NoStop}%
\bibitem [{\citenamefont {Sacha}\ \emph {et~al.}(2009)\citenamefont {Sacha},
  \citenamefont {M\"uller}, \citenamefont {Delande},\ and\ \citenamefont
  {Zakrzewski}}]{Sacha09}%
  \BibitemOpen
  \bibfield  {author} {\bibinfo {author} {\bibfnamefont {Krzysztof}\
  \bibnamefont {Sacha}}, \bibinfo {author} {\bibfnamefont {Cord~A.}\
  \bibnamefont {M\"uller}}, \bibinfo {author} {\bibfnamefont {Dominique}\
  \bibnamefont {Delande}}, \ and\ \bibinfo {author} {\bibfnamefont {Jakub}\
  \bibnamefont {Zakrzewski}},\ }\bibfield  {title} {\enquote {\bibinfo {title}
  {Anderson localization of solitons},}\ }\href {\doibase
  10.1103/PhysRevLett.103.210402} {\bibfield  {journal} {\bibinfo  {journal}
  {Phys. Rev. Lett.}\ }\textbf {\bibinfo {volume} {103}},\ \bibinfo {pages}
  {210402} (\bibinfo {year} {2009})}\BibitemShut {NoStop}%
\bibitem [{\citenamefont {Delande}\ \emph {et~al.}(2013)\citenamefont
  {Delande}, \citenamefont {Sacha}, \citenamefont {P{\l}odzie{\'{n}}},
  \citenamefont {Avazbaev},\ and\ \citenamefont
  {Zakrzewski}}]{delande2013manybody}%
  \BibitemOpen
  \bibfield  {author} {\bibinfo {author} {\bibfnamefont {Dominique}\
  \bibnamefont {Delande}}, \bibinfo {author} {\bibfnamefont {Krzysztof}\
  \bibnamefont {Sacha}}, \bibinfo {author} {\bibfnamefont {Marcin}\
  \bibnamefont {P{\l}odzie{\'{n}}}}, \bibinfo {author} {\bibfnamefont
  {Sanat~K}\ \bibnamefont {Avazbaev}}, \ and\ \bibinfo {author} {\bibfnamefont
  {Jakub}\ \bibnamefont {Zakrzewski}},\ }\bibfield  {title} {\enquote {\bibinfo
  {title} {Many-body {A}nderson localization in one-dimensional systems},}\
  }\href {\doibase 10.1088/1367-2630/15/4/045021} {\bibfield  {journal}
  {\bibinfo  {journal} {New Journal of Physics}\ }\textbf {\bibinfo {volume}
  {15}},\ \bibinfo {pages} {045021} (\bibinfo {year} {2013})}\BibitemShut
  {NoStop}%
\bibitem [{\citenamefont {Pitaevskii}\ and\ \citenamefont
  {Stringari}(2016)}]{pitaevskii2016bose}%
  \BibitemOpen
  \bibfield  {author} {\bibinfo {author} {\bibfnamefont {Lev}\ \bibnamefont
  {Pitaevskii}}\ and\ \bibinfo {author} {\bibfnamefont {Sandro}\ \bibnamefont
  {Stringari}},\ }\href@noop {} {\emph {\bibinfo {title} {{B}ose-{E}instein
  condensation and superfluidity}}},\ Vol.\ \bibinfo {volume} {164}\ (\bibinfo
  {publisher} {Oxford University Press},\ \bibinfo {year} {2016})\BibitemShut
  {NoStop}%
\bibitem [{\citenamefont {Akkermans}\ and\ \citenamefont
  {Montambaux}(2007)}]{akkermans2007mesoscopic}%
  \BibitemOpen
  \bibfield  {author} {\bibinfo {author} {\bibfnamefont {Eric}\ \bibnamefont
  {Akkermans}}\ and\ \bibinfo {author} {\bibfnamefont {Gilles}\ \bibnamefont
  {Montambaux}},\ }\href@noop {} {\emph {\bibinfo {title} {Mesoscopic physics
  of electrons and photons}}}\ (\bibinfo  {publisher} {Cambridge university
  press},\ \bibinfo {year} {2007})\BibitemShut {NoStop}%
\bibitem [{\citenamefont {Mikhailovich}\ \emph {et~al.}(1988)\citenamefont
  {Mikhailovich}, \citenamefont {Gredeskul},\ and\ \citenamefont
  {Pastur}}]{lifshits1988introduction}%
  \BibitemOpen
  \bibfield  {author} {\bibinfo {author} {\bibfnamefont {Lifshits~Il'ya}\
  \bibnamefont {Mikhailovich}}, \bibinfo {author} {\bibfnamefont
  {Sergeyi~Andreevich}\ \bibnamefont {Gredeskul}}, \ and\ \bibinfo {author}
  {\bibfnamefont {Leonid~Andreevich}\ \bibnamefont {Pastur}},\ }\href@noop {}
  {\emph {\bibinfo {title} {Introduction to the theory of disordered
  systems}}}\ (\bibinfo  {publisher} {Wiley-Interscience},\ \bibinfo {year}
  {1988})\BibitemShut {NoStop}%
\bibitem [{\citenamefont {Berezinskii}(1974)}]{berezinskii1974kinetics}%
  \BibitemOpen
  \bibfield  {author} {\bibinfo {author} {\bibfnamefont {VL}~\bibnamefont
  {Berezinskii}},\ }\bibfield  {title} {\enquote {\bibinfo {title} {Kinetics of
  a quantum particle in a one-dimensional random potential},}\ }\href
  {http://www.jetp.ac.ru/cgi-bin/e/index/e/38/3/p620?a=list} {\bibfield
  {journal} {\bibinfo  {journal} {Soviet Journal of Experimental and
  Theoretical Physics}\ }\textbf {\bibinfo {volume} {38}},\ \bibinfo {pages}
  {620} (\bibinfo {year} {1974})}\BibitemShut {NoStop}%
\bibitem [{Note1()}]{Note1}%
  \BibitemOpen
  \bibinfo {note} {Note that there is no constraint on the size of the wave
  packet $\sigma $ compared to the mean free path $\ell $. The expression (\ref
  {eq:x_vs_t}) does not depend on the ratio of the these two length
  scales.}\BibitemShut {Stop}%
\bibitem [{\citenamefont {Cherroret}\ and\ \citenamefont
  {Wellens}(2011)}]{cherroret2011fokker}%
  \BibitemOpen
  \bibfield  {author} {\bibinfo {author} {\bibfnamefont {Nicolas}\ \bibnamefont
  {Cherroret}}\ and\ \bibinfo {author} {\bibfnamefont {Thomas}\ \bibnamefont
  {Wellens}},\ }\bibfield  {title} {\enquote {\bibinfo {title}
  {{F}okker-{P}lanck equation for transport of wave packets in nonlinear
  disordered media},}\ }\href {\doibase 10.1103/PhysRevE.84.021114} {\bibfield
  {journal} {\bibinfo  {journal} {Physical Review E}\ }\textbf {\bibinfo
  {volume} {84}},\ \bibinfo {pages} {021114} (\bibinfo {year}
  {2011})}\BibitemShut {NoStop}%
\bibitem [{\citenamefont {Geiger}\ \emph {et~al.}(2013)\citenamefont {Geiger},
  \citenamefont {Buchleitner},\ and\ \citenamefont
  {Wellens}}]{geiger2013microscopic}%
  \BibitemOpen
  \bibfield  {author} {\bibinfo {author} {\bibfnamefont {Tobias}\ \bibnamefont
  {Geiger}}, \bibinfo {author} {\bibfnamefont {Andreas}\ \bibnamefont
  {Buchleitner}}, \ and\ \bibinfo {author} {\bibfnamefont {Thomas}\
  \bibnamefont {Wellens}},\ }\bibfield  {title} {\enquote {\bibinfo {title}
  {Microscopic scattering theory for interacting bosons in weak random
  potentials},}\ }\href {\doibase 10.1088/1367-2630/15/11/115015} {\bibfield
  {journal} {\bibinfo  {journal} {New Journal of Physics}\ }\textbf {\bibinfo
  {volume} {15}},\ \bibinfo {pages} {115015} (\bibinfo {year}
  {2013})}\BibitemShut {NoStop}%
\bibitem [{\citenamefont {Wellens}(2017-2019)}]{wellens2019private}%
  \BibitemOpen
  \bibfield  {author} {\bibinfo {author} {\bibfnamefont {Thomas}\ \bibnamefont
  {Wellens}},\ }\href@noop {} {}\bibinfo {howpublished} {private communication}
  (\bibinfo {year} {2017-2019})\BibitemShut {NoStop}%
\bibitem [{\citenamefont {Tal-Ezer}\ and\ \citenamefont
  {Kosloff}(1984)}]{Tal-Ezer84}%
  \BibitemOpen
  \bibfield  {author} {\bibinfo {author} {\bibfnamefont {H.}~\bibnamefont
  {Tal-Ezer}}\ and\ \bibinfo {author} {\bibfnamefont {R.}~\bibnamefont
  {Kosloff}},\ }\href@noop {} {\bibfield  {journal} {\bibinfo  {journal} {J.
  Chem. Phys.}\ }\textbf {\bibinfo {volume} {81}},\ \bibinfo {pages} {3967}
  (\bibinfo {year} {1984})}\BibitemShut {NoStop}%
\bibitem [{\citenamefont {Leforestier}\ \emph {et~al.}(1991)\citenamefont
  {Leforestier}, \citenamefont {Bisseling}, \citenamefont {Cerjan},
  \citenamefont {Feit}, \citenamefont {Friesner}, \citenamefont {Guldberg},
  \citenamefont {Hammerich}, \citenamefont {Jolicard}, \citenamefont
  {Karrlein}, \citenamefont {Meyer}, \citenamefont {Lipkin}, \citenamefont
  {Roncero},\ and\ \citenamefont {Kosloff}}]{Cheby91}%
  \BibitemOpen
  \bibfield  {author} {\bibinfo {author} {\bibfnamefont {C.}~\bibnamefont
  {Leforestier}}, \bibinfo {author} {\bibfnamefont {R.H.}\ \bibnamefont
  {Bisseling}}, \bibinfo {author} {\bibfnamefont {C.}~\bibnamefont {Cerjan}},
  \bibinfo {author} {\bibfnamefont {M.D.}\ \bibnamefont {Feit}}, \bibinfo
  {author} {\bibfnamefont {R.}~\bibnamefont {Friesner}}, \bibinfo {author}
  {\bibfnamefont {A.}~\bibnamefont {Guldberg}}, \bibinfo {author}
  {\bibfnamefont {A.}~\bibnamefont {Hammerich}}, \bibinfo {author}
  {\bibfnamefont {G.}~\bibnamefont {Jolicard}}, \bibinfo {author}
  {\bibfnamefont {W.}~\bibnamefont {Karrlein}}, \bibinfo {author}
  {\bibfnamefont {H.-D.}\ \bibnamefont {Meyer}}, \bibinfo {author}
  {\bibfnamefont {N.}~\bibnamefont {Lipkin}}, \bibinfo {author} {\bibfnamefont
  {O.}~\bibnamefont {Roncero}}, \ and\ \bibinfo {author} {\bibfnamefont
  {R.}~\bibnamefont {Kosloff}},\ }\bibfield  {title} {\enquote {\bibinfo
  {title} {A comparison of different propagation schemes for the time dependent
  schrödinger equation},}\ }\href {\doibase
  https://doi.org/10.1016/0021-9991(91)90137-A} {\bibfield  {journal} {\bibinfo
   {journal} {J. Comput. Phys.}\ }\textbf {\bibinfo {volume} {94}},\ \bibinfo
  {pages} {59} (\bibinfo {year} {1991})}\BibitemShut {NoStop}%
\bibitem [{\citenamefont {Roche}\ and\ \citenamefont
  {Mayou}(1997)}]{roche1997conductivity}%
  \BibitemOpen
  \bibfield  {author} {\bibinfo {author} {\bibfnamefont {S.}~\bibnamefont
  {Roche}}\ and\ \bibinfo {author} {\bibfnamefont {D.}~\bibnamefont {Mayou}},\
  }\bibfield  {title} {\enquote {\bibinfo {title} {Conductivity of
  quasiperiodic systems: A numerical study},}\ }\href {\doibase
  10.1103/PhysRevLett.79.2518} {\bibfield  {journal} {\bibinfo  {journal}
  {Phys. Rev. Lett.}\ }\textbf {\bibinfo {volume} {79}},\ \bibinfo {pages}
  {2518--2521} (\bibinfo {year} {1997})}\BibitemShut {NoStop}%
\bibitem [{\citenamefont {Fehske}\ \emph {et~al.}(2009)\citenamefont {Fehske},
  \citenamefont {Schleede}, \citenamefont {Schubert}, \citenamefont {Wellein},
  \citenamefont {Filinov},\ and\ \citenamefont {Bishop}}]{fehske2009numerical}%
  \BibitemOpen
  \bibfield  {author} {\bibinfo {author} {\bibfnamefont {Holger}\ \bibnamefont
  {Fehske}}, \bibinfo {author} {\bibfnamefont {Jens}\ \bibnamefont {Schleede}},
  \bibinfo {author} {\bibfnamefont {Gerald}\ \bibnamefont {Schubert}}, \bibinfo
  {author} {\bibfnamefont {Gerhard}\ \bibnamefont {Wellein}}, \bibinfo {author}
  {\bibfnamefont {Vladimir~S.}\ \bibnamefont {Filinov}}, \ and\ \bibinfo
  {author} {\bibfnamefont {Alan~R.}\ \bibnamefont {Bishop}},\ }\bibfield
  {title} {\enquote {\bibinfo {title} {Numerical approaches to time evolution
  of complex quantum systems},}\ }\href {\doibase
  https://doi.org/10.1016/j.physleta.2009.04.022} {\bibfield  {journal}
  {\bibinfo  {journal} {Physics Letters A}\ }\textbf {\bibinfo {volume}
  {373}},\ \bibinfo {pages} {2182 -- 2188} (\bibinfo {year}
  {2009})}\BibitemShut {NoStop}%
\bibitem [{Note2()}]{Note2}%
  \BibitemOpen
  \bibinfo {note} {A preliminary study of this partial destruction of the
  boomerang effect can be found on the arXiv preprint~\cite {prat2017anderson}.
  However, these preliminary results are not part of the published paper~\cite
  {prat2019quantumboomerang}}\BibitemShut {NoStop}%
\bibitem [{\citenamefont {Tavora}\ \emph {et~al.}(2014)\citenamefont {Tavora},
  \citenamefont {Rosch},\ and\ \citenamefont {Mitra}}]{Tavora2014}%
  \BibitemOpen
  \bibfield  {author} {\bibinfo {author} {\bibfnamefont {Marco}\ \bibnamefont
  {Tavora}}, \bibinfo {author} {\bibfnamefont {Achim}\ \bibnamefont {Rosch}}, \
  and\ \bibinfo {author} {\bibfnamefont {Aditi}\ \bibnamefont {Mitra}},\
  }\bibfield  {title} {\enquote {\bibinfo {title} {Quench dynamics of
  one-dimensional interacting bosons in a disordered potential: Elastic
  dephasing and critical speeding-up of thermalization},}\ }\href {\doibase
  10.1103/PhysRevLett.113.010601} {\bibfield  {journal} {\bibinfo  {journal}
  {Phys. Rev. Lett.}\ }\textbf {\bibinfo {volume} {113}},\ \bibinfo {pages}
  {010601} (\bibinfo {year} {2014})}\BibitemShut {NoStop}%
\bibitem [{\citenamefont {Cherroret}\ \emph {et~al.}(2015)\citenamefont
  {Cherroret}, \citenamefont {Karpiuk}, \citenamefont {Gr\'emaud},\ and\
  \citenamefont {Miniatura}}]{Miniatura15}%
  \BibitemOpen
  \bibfield  {author} {\bibinfo {author} {\bibfnamefont {Nicolas}\ \bibnamefont
  {Cherroret}}, \bibinfo {author} {\bibfnamefont {Tomasz}\ \bibnamefont
  {Karpiuk}}, \bibinfo {author} {\bibfnamefont {Beno\^{\i}t}\ \bibnamefont
  {Gr\'emaud}}, \ and\ \bibinfo {author} {\bibfnamefont {Christian}\
  \bibnamefont {Miniatura}},\ }\bibfield  {title} {\enquote {\bibinfo {title}
  {Thermalization of matter waves in speckle potentials},}\ }\href {\doibase
  10.1103/PhysRevA.92.063614} {\bibfield  {journal} {\bibinfo  {journal} {Phys.
  Rev. A}\ }\textbf {\bibinfo {volume} {92}},\ \bibinfo {pages} {063614}
  (\bibinfo {year} {2015})}\BibitemShut {NoStop}%
\bibitem [{\citenamefont {Abrahams}\ \emph {et~al.}(1979)\citenamefont
  {Abrahams}, \citenamefont {Anderson}, \citenamefont {Licciardello},\ and\
  \citenamefont {Ramakrishnan}}]{Abrahams79}%
  \BibitemOpen
  \bibfield  {author} {\bibinfo {author} {\bibfnamefont {E.}~\bibnamefont
  {Abrahams}}, \bibinfo {author} {\bibfnamefont {P.~W.}\ \bibnamefont
  {Anderson}}, \bibinfo {author} {\bibfnamefont {D.~C.}\ \bibnamefont
  {Licciardello}}, \ and\ \bibinfo {author} {\bibfnamefont {T.~V.}\
  \bibnamefont {Ramakrishnan}},\ }\bibfield  {title} {\enquote {\bibinfo
  {title} {Scaling theory of localization: Absence of quantum diffusion in two
  dimensions},}\ }\href {\doibase 10.1103/PhysRevLett.42.673} {\bibfield
  {journal} {\bibinfo  {journal} {Phys. Rev. Lett.}\ }\textbf {\bibinfo
  {volume} {42}},\ \bibinfo {pages} {673--676} (\bibinfo {year}
  {1979})}\BibitemShut {NoStop}%
\bibitem [{\citenamefont {Goodman}(2007)}]{goodman2007speckle}%
  \BibitemOpen
  \bibfield  {author} {\bibinfo {author} {\bibfnamefont {Joseph~W}\
  \bibnamefont {Goodman}},\ }\href@noop {} {\emph {\bibinfo {title} {Speckle
  phenomena in optics: theory and applications}}}\ (\bibinfo  {publisher}
  {Roberts and Company Publishers, Greenwood Village (Co)},\ \bibinfo {year}
  {2007})\BibitemShut {NoStop}%
\bibitem [{Note3()}]{Note3}%
  \BibitemOpen
  \bibinfo {note} {The averaging must in principle be performed over many
  disorder realizations, while the interaction energy may be different for each
  disorder realization. In practice, a spatial averaging over the wave packet
  size is equivalent, provided the wave packet contains many speckle grains,
  which is the case if $\sigma \gg 1/k_0$.}\BibitemShut {Stop}%
\bibitem [{\citenamefont {Prat}\ \emph {et~al.}(2017)\citenamefont {Prat},
  \citenamefont {Delande},\ and\ \citenamefont {Cherroret}}]{prat2017anderson}%
  \BibitemOpen
  \bibfield  {author} {\bibinfo {author} {\bibfnamefont {Tony}\ \bibnamefont
  {Prat}}, \bibinfo {author} {\bibfnamefont {Dominique}\ \bibnamefont
  {Delande}}, \ and\ \bibinfo {author} {\bibfnamefont {Nicolas}\ \bibnamefont
  {Cherroret}},\ }\bibfield  {title} {\enquote {\bibinfo {title} {When
  {A}nderson localization makes quantum particles move backward},}\ }\href
  {https://arxiv.org/abs/1704.05241v1} {\bibfield  {journal} {\bibinfo
  {journal} {arXiv preprint arXiv:1704.05241v1}\ } (\bibinfo {year}
  {2017})}\BibitemShut {NoStop}%
\end{thebibliography}%

\end{document}